\documentclass[10pt]{spora}

\usepackage{amsmath} 
\usepackage{bm}
\usepackage{multirow}
\usepackage{enumitem}

\title[Effective dose fractionation of radiotherapy ]{Effective dose fractionation schemes of radiotherapy for prostate cancer}

\addauthor{Jose Alvarez}{Alvarez}{1}
\addlocation{1}{University of California, Riverside, USA}

\addauthor{Kathleen M. Storey}{Storey}{2}
\addlocation{2}{Lafayette College, USA}

\addauthor{Pavitra Kannan}{Kannan}{3}

\addlocation{3}{Karolinska Institute, Stockholm, Sweden}
\addauthor*{Heyrim Cho}{Cho}{1}
\address{Prof.~Heyrim Cho,
  Dept.\ of Mathematics,
  University of California, Riverside,
  900 University Ave, Riverside, CA 92521,
  USA
}
\email{heyrim.cho@ucr.edu}


\abstract{%
Radiation therapy has remained as one of the main cancer treatment modalities and a highly cost-effective single modality treatment of cancer care. Typical regimens for fractionated external beam radiotherapy comprise a constant dose administered on weekdays, and no radiation on weekends. However, every patient has a tumor with distinct properties depending on intra-tumor heterogeneity, aggressiveness, and interactive properties with other cells that may make it more resistant or sensitive to radiation treatment. Accordingly, the concept of personalized cancer treatment is emerging to specialize each patient treatment case to the unique properties of the tumor. In this paper, we examine adaptive radiation treatment strategies for heterogeneous tumors {using a dynamical system model} that consist of radiation-resistant and parental cell populations with unique interactive properties. We study different adaptive dosage strategies for PC3 and DU145 prostate cancer cell lines. We show that stronger doses of radiation given in longer time intervals, while keeping the overall dosage the same, reduce final tumor volume by more than half in PC3 cell lines, but by only five percent in DU145 cell lines. In addition, we tested an adaptive dosing schedule by administering a stronger dosage on Friday to compensate for the treatment-off period during the weekend, which was effective in decreasing the final tumor volume of both cell lines. This result creates interesting possibilities for new radiotherapy strategies at clinics that cannot provide treatment on weekends. Finally, we propose a dosage plan incorporating our findings. {It is imperative to clarify that this was a simulation-based study using ordinary differential equations that has yet to be verified experimentally and clinically. However, we hope to further study the factors among the complex nature of the tumor that close the gap between our simulation result and reality.} 
}

\begin{document}

\maketitle

%%%%%%%%%%%%%%%%%%%%%%%%%%%%%%%%%%%%%%%%
%%%%%%%%%%%%%%%%%%%%%%%%%%%%%%%%%%%%%%%%

\section{Introduction}

% [Radiotherapy for cancer]
Radiotherapy is one of the predominant cancer treatment modalities. Approximately 50\% of all cancer patients receive radiotherapy treatments during the course of their illness  \cite{Delaney2005,Begg2011}, and it is estimated that radiation therapy contributes around 40\% towards curative treatment \cite{Barnett2009}. It is also a highly cost effective single modality treatment, accounting for only about 5\% of the total cost of cancer care \cite{Ringborg2003}. Furthermore, advances in imaging techniques, computerized treatment planning systems, radiation treatment machines with improved X-ray production and treatment delivery, as well as improved understanding of the radiobiology are all increasing the impact and importance of radiotherapy \cite{Bernier2004}.

% [Dose fractionation]
In clinics, the dosing schedule of radiotherapy is standardized using daily doses of 1.8 to 2.0 for 39--45 fractions \cite{Vanneste2016}. This is called fractionation, where the total dosage is divided into  smaller doses that are given over a period of one to two months. The specific dose for each patient depends on the location and severity of the tumor, as well as the radiation-induced toxicity of normal tissues surrounding the tumor. 
Various studies focus on dose alteration to improve radiotherapy outcome, including hypofractionation and hyperfractionation \cite{Pollack}. Hypofractionation is a treatment regimen that uses higher doses of radiation in fewer visits to lower the effects of accelerated tumor growth that typically occurs during the later stages of radiotherapy. 
On the other hand, hyperfractionation is a strategy dividing the same total dose into more frequent deliveries, for instance, radiation doses given more than once a day. 
A recent study in \cite{Zaorsky2015} shows that increasing the total dosage from 140 to 200 Gy for prostate cancer is associated with improved outcome, but doses above 200 Gy did not result in additional clinical benefit. Moreover, population-based research revealed an association between overall survival of prostate cancer patients in doses over 75.6 Gy \cite{Vanneste2016}. However, the dosage schemes that can be tested in clinical trials are very limited. Simulation using in-silico models can help address this limitation to test various dosing schedules without the concern of toxicity to patients.

% [Math modeling of cancer]
A large number of mathematical and computational models have been developed to study tumor growth and cancer treatments, including  differential equations \cite{Murphy2016,Koziol2020,Cho2017a}, multiphase models based on mixture theory \cite{Byrne2010,Sunassee2019} or phase field theory \cite{Shannon}, and multiscale models that couple subcellular, cellular, and tissue scale phenomena \cite{Kannan2019}. 
The  availability  of  detailed  information  about  tumors  has undoubtedly stimulated this field to more complex models, although as the model complexity increases, it becomes more challenging to uniquely identify the model parameter values \cite{Cho2020}. {In particular, often in the clinical setting, it is not possible to collect appropriate amounts of patients’ data to calibrate complex models. Hence, simple models such as ordinary differential equations (ODEs) are commonly used when dealing with clinical data. See \cite{Enderling2019,cho2020bayesian,Mohammad2021} for recent literature using ODE models to calibrate data of radiotherapy treated tumors as in our study.
}

% [Math modeling of radiotherapy]
Radiotherapy also has a long history of mathematical modeling. Radiation dosing is typically modeled using the linear-quadratic (L-Q) model \cite{Thames1982,Fowler1989}. Several recent studies have applied the L-Q model to patient-specific data, in an effort to evaluate and predict individual responses to radiotherapy \cite{Rockne2010,Corwin2013,Sunassee2019,Enderling2019}. Logistic type of radiotherapy response has been proposed with a concept of proliferation saturation index, defined as the ratio of tumor volume to the host-influenced tumor carrying capacity, that correlates inversely with radiotherapy response \cite{Prokopiou2015,Poleszczuk2018}. 
Other radiotherapy models, including those assuming a dynamic carrying capacity, have been developed to more accurately calibrate and predict individual patient response to radiotherapy \cite{Mohammad2021}. 
%{In our work, we will employ the most commonly used radiotherapy model, that is, the L-Q model, since our focus is not on developing a new radiotherapy model, but to study adaptive radiotherapy treatment to heterogeneous prostate cancer.} 

% [Aim of work]
Here, we investigate the effect of different radiotherapy regimens on the growth of two types of heterogeneous prostate tumors comprising radiation-sensitive {(or parental) populations} and resistant populations that interact with one another. {Although adaptive radiotherapy and its potential clinical benefits have been proposed in the clinical community since 1997 \cite{yan1997}, it has not been a common practice in the clinics until now \cite{brock2019}. However, recent improvements in imaging technologies along with mathematical modeling have been proclaimed to allow us new opportunities toward patient-specific adaptive radiation therapy \cite{Enderling2019}. Our work is in accordance with this idea of using mathematical models to guide radiation dose planning, which has been done for various types of tumors, such as glioblastomas \cite{leder2014} and lung cancer \cite{hong2021lung}. Our study specifically considers two types of prostate cancer \cite{Kannan2019}.} The paper is structured as follows. Section 2 summarizes the mathematical model that is used to describe the growth of heterogeneous cancer and its response to radiotherapy with biological interpretations of the model parameters.  We also describe how radiotherapy fractionation treatments are incorporated into the model in order to properly simulate cell death due to radiotherapy. In Section 3 we explore two different scenarios on dose fractionation. Section 3.1 studies a constant dosage treatment, while we change the dosage and time interval between the administration of radiation. The overall dosage is kept constant. In section 3.2, we study a strategy to overcome the clinical radiotherapy schedule with no radiotherapy treatments being administered on weekends. We compare the strategy of changing the dosage to be more heavily concentrated on Fridays at the end of the week to a dosage that is spread out evenly. Section 4 concludes our paper and summarizes our key results. In addition, it includes our proposed fractionation scheme that incorporates our findings into a possible dosage plan.

%%%%%%%%%%%%%%%%%%%%%%%%%%%%%%%%%%%%%%%%
%%%%%%%%%%%%%%%%%%%%%%%%%%%%%%%%%%%%%%%%

\section{Mathematical model of cancer growth and radiotherapy}

\begin{table*}[t] \centering
	\begin{tabular}{|c|c|c|c|c|} \hline 
	 & Biological meaning \\ \hline 
	$V_c(t)$ & Volume of the {control (radiation-sensitive/ parental) tumor} population at time $t$ 
	\\ 
	$V_r(t)$ & Volume of the radiation-resistant tumor population at time $t$\\ 
	$p_c$ & Rate of growth of the control population \\
	$p_r$ & Rate of growth of the radio-resistant population \\
	$K_c$ & Carrying capacity of the control population  \\
	$K_r$ & Carrying capacity of the radio-resistant population \\ 
	$\lambda_c$ & Effect of the control cell population on the radio-resistant cells \\
	$\lambda_r$ & Effect of the radio-resistant cell population on the control cells \\
 $\alpha_c$, $\beta_c$  & Radio-sensitivity parameter of control cells \\
 $\alpha_r$, $\beta_r$  & Radio-sensitivity parameter of radio-resistant cells \\
 \hline 
	\end{tabular}
	\caption{Model parameters and their biological interpretation. }
	\label{Tbl:param1}
\end{table*}

{The Lotka-Volterra model is one of the typical approaches to describe the interactions between multiple types of cancer cells \cite{gatenby1995,zhang2017Gatenby,freischel2021Gatenby,Liu2007,Kannan2019,Kuang2018}.} 
We also employ the Lotka-Volterra model to describe the growth of mixtures consisting of  parental and radioresistant tumor cell populations. The equation tracks the dynamics of $V_c(t)$, which is the volume of the parental tumor population, or the {“control”, in other words, the radiation-sensitive} population, and $V_r(t)$, which is the volume of the radiation-resistant tumor population.
\begin{equation}
    \begin{split}
\frac{dV_c}{dt} &= p_c V_c \left(1-\frac{V_c}{K_c} -\lambda_r \frac{V_r}{K_c}\right),  \\
\frac{dV_r}{dt} &= p_r V_r \left(1-\frac{V_r}{K_r} -\lambda_c \frac{V_c}{K_r} \right) . 
    \end{split}
\label{eqn:model}
\end{equation}
The parameters used in the Lotka-Volterra model each have their own biological function. {The rate of growth of the control population is $p_c$, and the rate of growth of the radio-resistant population is $p_r$.} $K_c$ is the carrying capacity of the control population, the maximum volume to which the {radio-sensitive or parental population} is limited to, and $K_r$ is the carrying capacity of the radioresistant population. In addition, $\lambda_c$ and $\lambda_r$ model the interaction between the two populations{, where} $\lambda_c$ describes the effect that the parental cell population has on radioresistant cells, and $\lambda_r$ describes the effect that the radioresistant cell population has on the parental cells. We note that $-1 \leq \lambda_c,\,\lambda_r \leq 1$. 
{The signs of the interaction parameters $\lambda_c$ and $\lambda_r$ need not be equivalent. A positive $\lambda_c$ parameter represents a detrimental effect on $V_r$, or the volume of the resistant tumor population, and a negative $\lambda_c$ parameter represents a beneficial effect on $V_r$. The same is true for the effects of $\lambda_r$ on $V_c$. If both interaction parameters are positive then this represents a competitive interaction between the two tumor populations, and if both are negative then this represents a mutualistic interaction. One positive and one negative interaction parameters signify that one tumor population exerts a positive effect on the other while the latter exerts a negative effect on the former tumor population, a so-called antagonistic interaction. An interaction parameter of 0 represents no effect on the other tumor population.}
All model parameters and their biological interpretations are summarized in Table \ref{Tbl:param1}.

\begin{table}[!htb]
\begin{center}
\begin{tabular}[c]{|c|c|c|c|}
        \hline
        \multirow{2}{*}{{Parameter}} & \multicolumn{2}{|c|}{Cell line}  & \multirow{2}{*}{{Units}}  \\ \cline{2-3}  
  & PC3 & DU145 & \\ 
\hline 
 $p_c$ & 0.36 & 0.6 & {day$^{-1}$} \\
 $p_r$ &  0.48 & 0.36 & {day$^{-1}$} \\
 $K_c$ &  0.85 & 0.75 & {mm$^{3}$} \\
 $K_r$ & 2.0 & 1.4 & {mm$^{3}$}  \\
 $\lambda_c$ &  0.2 & 0.25 & {day$^{-1}$mm$^{-3}$} \\
 $\lambda_r$ &  0.0 & -0.5 & {day$^{-1}$mm$^{-3}$} \\
 $\alpha_c$ &  0.43 & 0.2843 & {Gy$^{-1}$} \\
 $\alpha_r$ &  0.3 & 0.23 & {Gy$^{-1}$} \\
 $\beta_c$ &  0.0407 & 0.0161 & {Gy$^{-2}$} \\
 $\beta_r$ &  0.0402 & 0.0124 & {Gy$^{-2}$} \\
\hline 
\end{tabular}
\caption{{A summary of the parameter values used in the simulation for prostate cancer cell lines, PC3 and DU145. The pre-treatment parameters $p, K$, and $\lambda$ are taken from \cite{Kannan2019}, provided in supplementary Table S1 and Figure 2d of \cite{Kannan2019}. The radiotherapy parameters $\alpha$ and $\beta$ are estimated from data in supplementary Figure S1b in \cite{Kannan2019}.}}
\label{table:params}
\end{center}
\end{table}

%\subsection{Radiotherapy}
We now explain how we incorporate treatment with radiotherapy in the cancer model Eq. \eqref{eqn:model}. We consider a typical tumor treatment regimen in which daily doses of $d$ Gy are administered Monday through Friday for 6 weeks. We use the linear-quadratic model \cite{Hall1994} to account for the effect of radiotherapy. This model assumes that the fraction of cells that survive exposure to a dose $d$ of radiotherapy is given by
\begin{equation}
\mbox{Survival fraction}\ \  = e^{-\alpha d-\beta d^2},
\end{equation}
where $\alpha$ and $\beta$ are tissue specific radiosensitivity parameters that model single and double strand breaks of the DNA \cite{Lea1942}.  We assume that the effect of radiotherapy is instantaneous, with the non-surviving cell fraction immediately removed when therapy is administered. In particular, we denote the dosing schedule by $u(t)$ as the following summation of indicator functions,
\begin{equation}
u(t) = \sum_{i=1}^N {\delta(t-t_i)},     
\end{equation}
where {$\delta(t)$ is a Dirac-delta function that is $\delta(0)=1$ and zero elsewhere. This makes $u(t)$ to be one only at $t_i$ (for $i=1, 2, \ldots, N$), the times at which radiotherapy is delivered.}
As we combine the radiotherapy model with the radiosensitivity cancer growth model, we have  
\begin{eqnarray}
\frac{dV_c}{dt} &=& p_c V_c \left(1-\frac{V_c}{K_c} -\lambda_r \frac{V_r}{K_c}\right), \nonumber \\ &\,& \qquad \quad   
-(1-e^{-\alpha_c d-\beta_c d^2}) \: u(t) V_c, \\
\frac{dV_r}{dt} &=& p_r V_r \left(1-\frac{V_r}{K_r} -\lambda_c \frac{V_c}{K_r} \right) \nonumber \\ &\,& \qquad \quad 
-\underbrace{(1-e^{-\alpha_r d-\beta_r d^2}) \: u(t) V_r}_{\mbox{cell death due to RT}}.
\label{eqn:twocomp_RT}
\end{eqnarray}
Here, $\alpha_c$ and $\beta_c$ are radiosensitivity parameters for the control cells, and $\alpha_r$ and $\beta_r$ are for the resistant cells. 
%
%%%%%%%%%%%%%%%%%%%%%%%%%%%%%%%%%%%%%%%%
%%%%%%%%%%%%%%%%%%%%%%%%%%%%%%%%%%%%%%%%

\section{Simulation of radiotherapy dosage fractionation }

{These simulations will investigate dosage strategies in two separate prostate cancer cell lines, PC3 and DU145. Important to note when discussing each cell lines' response to varying dosage strategies is that each cell line varies in their growth rate, response to radiation, carrying capacity, and in the interaction parameters between control and radio-resistant cell populations. For instance, regarding the interaction between the control and radio-resistant population, PC3 has competitive interaction, while the control population of DU145 has antagonistic relation to the radio-resistant population and only the control population benefits from coexistence. Also, regarding sensitivity to radiotherapy, PC3 is more sensitive to the treatment compared to DU145. See Table \ref{table:params} for the parameter values. 
}

\begin{figure}[htb!]
\centerline{  
        \includegraphics[width=9cm]{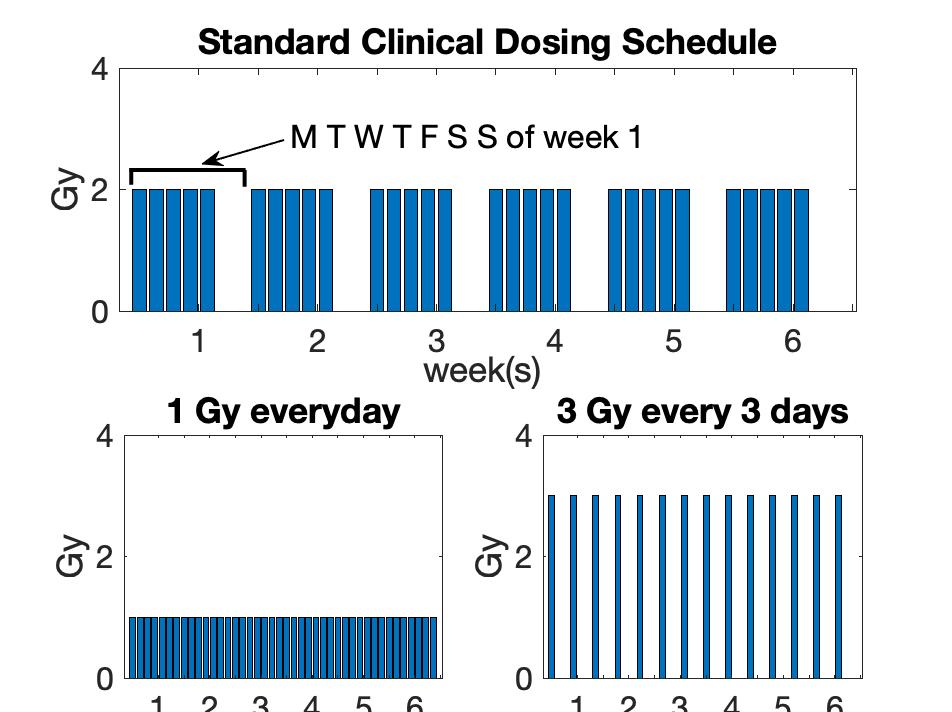} }
\centerline{  
        \includegraphics[width=9cm]{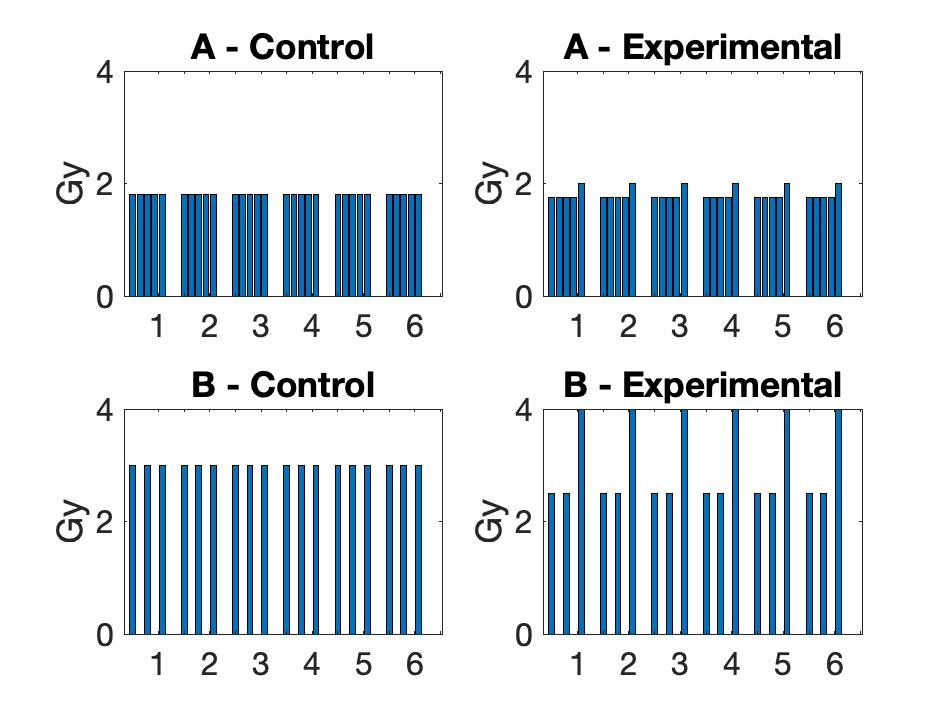} }
    \caption{ {Examples of radiotherapy dosage schedules tested in this study. The clinical standard dosage plan for prostate cancer is 2 Gy on weekdays, Monday to Friday, for 6 weeks (top). 
    The middle figures show examples of constant dosage schedules studied in section \ref{sec:simulation1}, e.g., 1 Gy every day plan (middle, left) and 3 Gy every 3 days plan (middle, right). 
     The bottom figures show the dosage schedules tested in section \ref{sec:simulation2}, e.g., Control schedules using constant dosage  (bottom, left) and Experimental schedules using stronger dosage on Friday  (bottom, right). } 
    }
    \label{fig:Dosageplan1}
\end{figure}

In our simulations, we consider a period of 6 weeks {of} radiotherapy treatment with an initial 2 week cancer growth period. 
{The dosage plans that we study are described in each section in detail, but see Figure \ref{fig:Dosageplan1} for some examples and a brief summary. }
The initial condition for the model is taken as {$V_c(0) = 0.5 \textrm{mm}^3$, and $ V_r(0)= 0.5 \textrm{mm}^3$}, assuming the parental and resistant tumor cell lines are initially in a 1:1 ratio. Although we test only for 1:1 ratio, we report on the control and resistant volume separately to examine the efficacy of treatment on each population. 

{In order to best measure the efficacy of radiation treatment, we decided to quantify two basic features of each simulation as a metric to determine the magnitude of influence each of our dosage strategies had on the growth of the simulated tumor population. Those are tumor volume and the tumor volume integrated in time, or area under the curve. While tumor volume is the most obvious metric to follow in order to measure the efficacy of simulated treatment, the area under the curve is also an important metric to follow because it represents the tumor volume integrated over the treatment period, which gives you an idea of the overall burden placed on the patient during that period of time.}

\subsection{Comparison of constant dosage strategies with different dose levels }
\label{sec:simulation1} 

\begin{figure}[htb!]
\centerline{  
        \includegraphics[width=4.2cm]{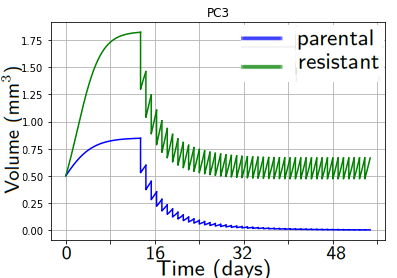}
        \includegraphics[width=4.2cm]{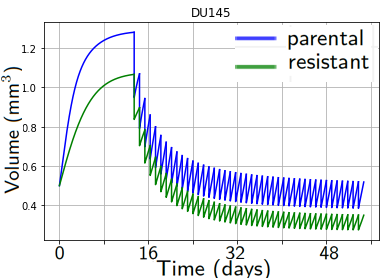}
        }
\centerline{  
        \includegraphics[width=4.2cm]{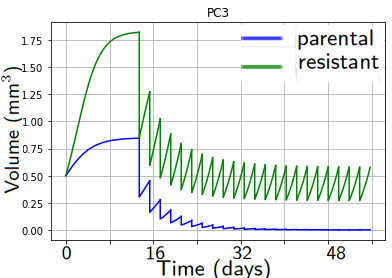}
        \includegraphics[width=4.2cm]{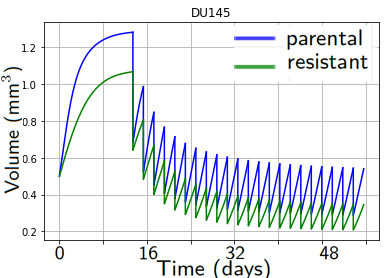}
        }
    \caption{Comparison of 1 Gy administered every 1 day (top) versus 2 Gy administered every 2 days (bottom) to a tumor {model initiated} as a 1:1 mixture of {parental} and resistant tumor cell populations. The results show the volumes for the individual {parental (blue)} and resistant {(green)} populations from the PC3 cell line (left) and DU145 cell line (right). When subjected to this change, a noticeable decrease in average tumor volume can be seen in the PC3 cell line when comparing the 1 Gy schedule and 2 Gy schedule. This change in average tumor size is not as noticeable in the DU145 cell line, as average tumor size remains unaffected, but a larger {variation of values} is noticed. See Tables \ref{TblA:PC3}-\ref{TblA:DU145} and Figure \ref{fig:A2} for further comparison. }
    \label{fig:A}
\end{figure}

\begin{figure}[htb!]
\centerline{  
        \includegraphics[width=4.2cm]{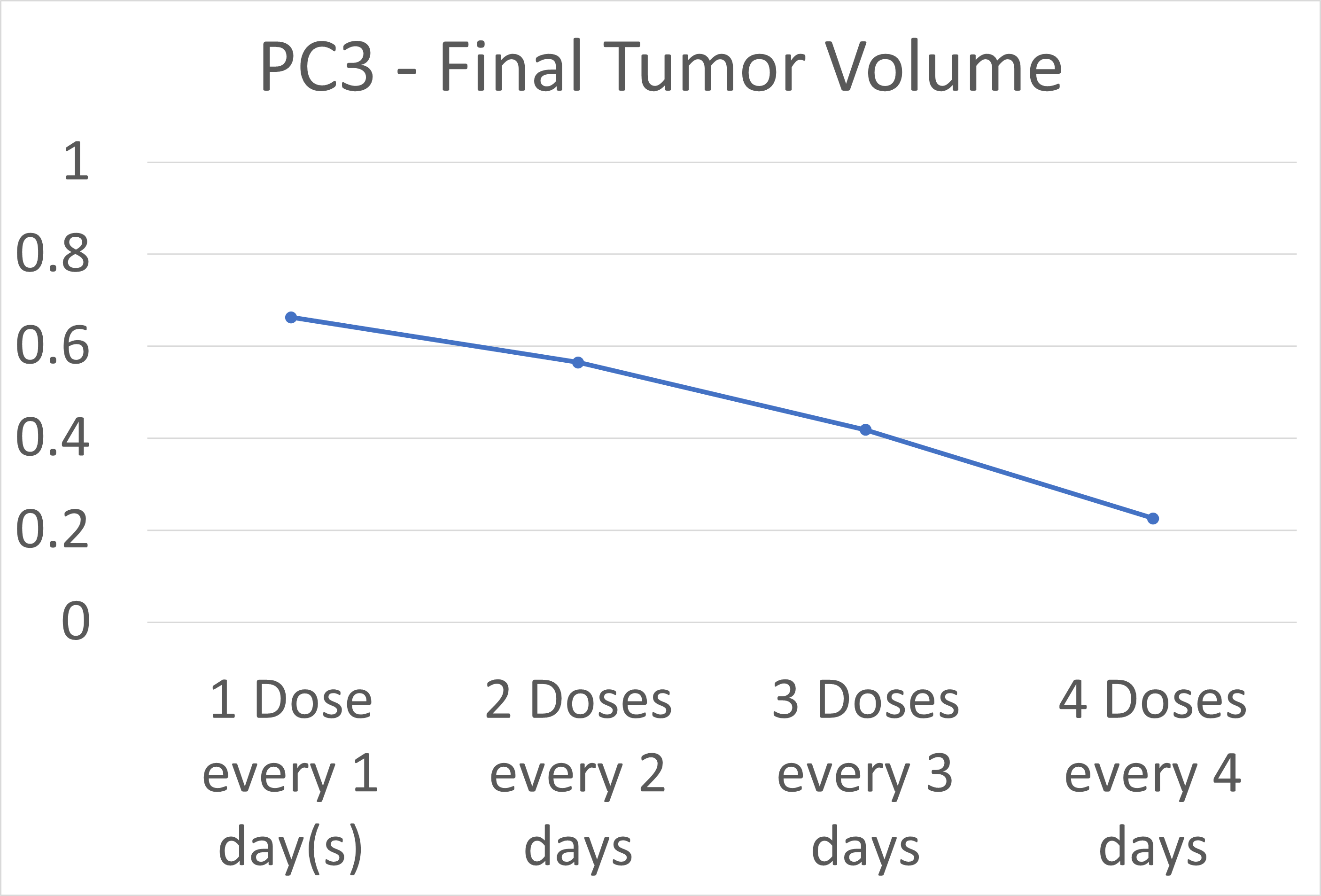}
        \includegraphics[width=4.4cm]{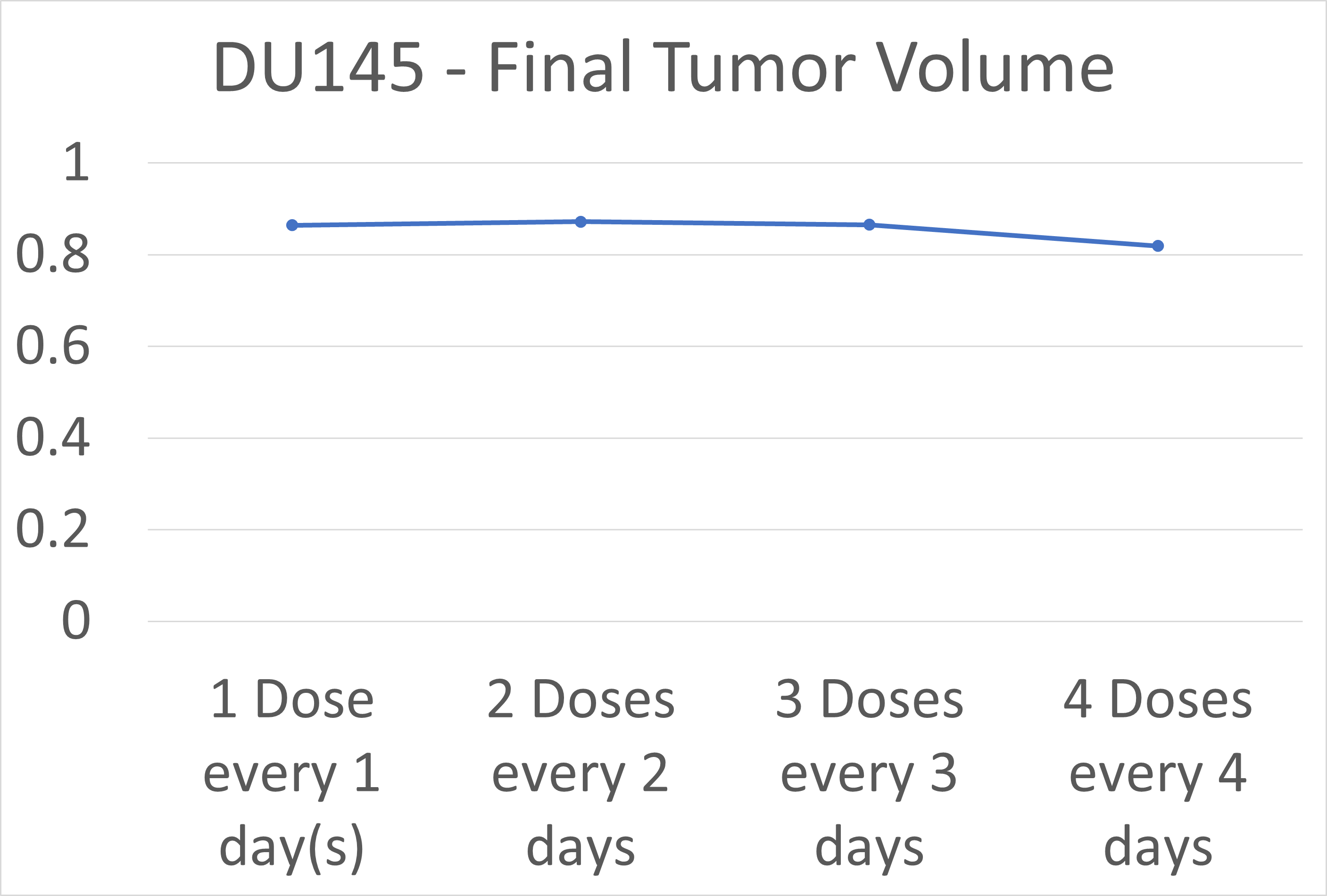}
        }
    \caption{{A line graph representing total final tumor volume for the PC3 cell line (left) based on values from Table \ref{TblA:PC3} as well as the DU145 cell line (right) based on values in Table \ref{TblA:DU145}. Increasing both dosage values and the time intervals in between dosages will not change the overall dosage, so results are dependent on hypofractionation and not an increased overall dosage. This strategy provides promising results for the PC3 cell line, but not so much for the DU145 cell line. }}
    \label{fig:A2}
\end{figure}

In this section, we study radiotherapy schedules in which the dosage value and time interval remain constant over the six-week treatment period. In particular, we compare different dosage levels, from 1 Gy to 4 Gy, while increasing the time interval between each administered dose for stronger dosages to keep the total dose at the end of 6 weeks constant, at 42 Gy in total. The case of 4 Gy has a lower total dosage due to the 6 week time constraint, but still continues the observed trend even at 40 Gy total. This strategy is called hypofractionation. 

In Tables \ref{TblA:PC3} and \ref{TblA:DU145}, we can compare the results generated by ramping up dosage values as well as increasing the time in between each administered dose. By organizing these events according to their respective dosage and time interval increase, we can show a clear trend in the data for the PC3 cell line, as shown in Table \ref{TblA:PC3}, in which a higher dose administered over a greater period of time gives better results than a lower dose administered over a shorter length of time. {For example, 2 Gy administered every 2 days yields better results than 1 Gy administered every 1 day.} As an example, in Figure \ref{fig:A}, we show the tumor trajectory for the case of comparing a single (1 Gy) dose administered every 1 day versus {a 2 Gy dose} administered every 2 days in both cell lines. {Figure \ref{fig:A} follows the basic parameters listed in Table 2. It is important to note that the PC3 and DU145 cell lines are defined by their own separate parameters and therefore present with varying changes after dosage administration. For example, the resistant population of tumor cells in the PC3 cell line have a higher carrying capacity, growth rate, and higher levels of radio-sensitivity than do resistant cells in the DU145 cell line. The same is true for parental cells in the DU145 cell line vs. parental cells in the PC3 cell line. This is why, despite it being seemingly counterintuitive, we see resistant cells end with a smaller volume than the more radiation-sensitive parental cells in the DU145 cell line. In this instance, one should consider the growth rate of the parental population and how it far exceeds that of the resistant cells.} 

{In Figure \ref{fig:A}} we can see significant decay for the PC3 cell line in both final tumor size and the area under the curve. However, results did not prove to be as effective for the DU145 cell line, as shown in Table \ref{TblA:DU145} and Figure \ref{fig:A2}, since only a slight decrease in the area under the curve is observable, and hardly any change can be noted from the final volume of the tumor. In particular, the parental DU145 cell line shows increasing final tumor volume as dosages become stronger. {Therefore, out of the two cell lines, it would be safe to assume that this dosage strategy proves effective only for the PC3 cell line and not the DU145 cell line. We will further investigate the underlying mechanism that makes the contrast between the two cell lines in the future.}

\begin{table*}[t] \centering
	\begin{tabular}{|c|c|c|c|c|c|} \hline 
	 & Dosage (Gy) & 1 & 2 & 3 & 4 \\ \hline
	 & Time interval (days) & 1 & 2 & 3 & 4 \\ \hline 
	 \multirow{3}{*}{Final Tumor Volume} & Parental & 0.0020287 & 0.00043197 & 0.000076458 & 0.0000069054\\ 
	 & Resistant & 0.66080 & 0.56423 & 0.41890 & 0.22569\\
	 & Total & 0.662823 & 0.56466 & 0.41808 & 0.22570\\ \hline 
	 \multirow{3}{*}{Area Under Curve} & Parental & 298.16 & 277.67 & 262.53 & 251.33 \\ & Resistant & 1046.4 & 911.99 & 791.61 & 694.12\\ & Total & 1344.5 & 1189.7 & 1054.1 & 945.45\\ \hline

\hline 
	\end{tabular}
	\caption{PC3 Cell line: Final tumor volume after treatment and area under the curve following an increase in time interval with stronger dosage, while maintaining a constant overall dosage (i.e., under different hypofractionation strategies). A steep decrease in final tumor volume was observed, as well as a decrease in the area under the curve.}
	\label{TblA:PC3}
\end{table*}

\begin{table*}[t] \centering
	\begin{tabular}{|c|c|c|c|c|c|} \hline 
	 & Dosage (Gy) & 1 & 2 & 3 & 4 \\ \hline
	 & Time interval (days) & 1 & 2 & 3 & 4 \\ \hline 
	 \multirow{3}{*}{Final Tumor Volume} & Parental & 0.5162 & 0.53188 & 0.53996 & 0.526363 \\ 
	 & Resistant & 0.34821 & 0.34030 & 0.32530 & 0.29257\\
	 & Total & 0.86444 & 0.87218 & 0.86526 & 0.81893\\ \hline 
	 \multirow{3}{*}{Area Under Curve} & Parental & 836.98 & 794.75 & 757.06 & 742.58 \\ & Resistant & 661.44 & 624.52 & 590.47 & 569.93 \\ & Total & 1498.4 & 1419.3 & 1347.5 & 1312.5\\ \hline

\hline 
	\end{tabular}
	\caption{DU145 Cell line: Final tumor volume after treatment and area under the curve following an increase in time interval with stronger dosage, while maintaining a constant overall dosage (i.e., under different hypofractionation strategies). While a decrease in the area under the curve can be noted, the final tumor volumes of parental and resistant populations yielded results with no obvious downward trend.}
	\label{TblA:DU145}
\end{table*} 

\subsection{Comparison of uniform dosage versus stronger dosage on Fridays }
\label{sec:simulation2} 

Next, we study various scenarios in which no radiation would be administered over the weekend, to more closely resemble a {real-life scenario in which the} radiotherapy clinic would be closed on the weekends, therefore unable to administer any radiation to a patient during this time frame. {This weekend time frame allows for unchecked tumor growth for approximately 3 days until the next dose of radiation is administered. The standard radiation dosage strategy for most radiotherapy clinics would be a consistent administration of 2 Gy of radiation every weekday, and 0 Gy on weekends. However, instead of administering a constant radiation dose of 2 Gy every weekday, we decided to allocate more radiation at the end of the week (on Friday) at the expense of lowering the radiation dose for all of the days prior. The idea is to maintain the same total amount of radiation at the end of each week so as to not compromise the total amount of radiation the patient receives, however instead of maintaining a consistent dosage throughout, we can modify the administration of radiation so that more of it is administered at the end of each week in order to combat weekend tumor growth.}

We considered {two} different scenarios ({Scenario A and Scenario B}) to test how {a simulated tumor model might shift} as we vary the {dosage values} and their respective time intervals. {We also provide a variation on either scenario (Scenario A$^\prime$ and Scenario B$^\prime$) in order to confirm the consistency of a stronger Friday dose by changing the Friday dosage value. } We compared the results from dosing every weekday (Scenarios A and {A$^\prime$}) versus dosing every Monday, Wednesday, and Friday ({Scenarios B and B$^\prime$}). However, we {integrated the aforementioned guiding principle in this investigation}, i.e.~stronger doses given at the end of each week {on Fridays}, to compensate for {the weekend} time intervals {spent} without radiotherapy. For example, in this investigation, we administered a {higher} dosage on Fridays than on {any} other weekday, and then we allowed the tumor to “re-grow” on the weekends to represent a lack of radiotherapy treatment given within that time frame. {We used “end-of-week" values as our main metric for determining the effectiveness of each treatment modification, and each of these values are measured every Friday after radiation administration.} Each scenario is as follows. {We remark that the dosage schedule of scenarios A and B are plotted as bar graph in the bottom of Figure \ref{fig:Dosageplan1}.}

\begin{itemize}

    \item {Scenario A involved the simulated administration of 1.8 Gy every day for five days to total 9 Gy at the end of every week as our control group. In our experimental group this weekly 9 Gy value was maintained while varying daily dosage values. The dosage value in our experimental group would increase from 1.75 Gy every day Monday-Thursday, to a slightly higher dose of 2 Gy every Friday. This sums to 9 Gy at the end of every week, same as the control group. Figures \ref{fig:table5} and \ref{fig:table6} depict the results of these changes in either of the PC3 or DU145 cell lines, which shows only a slight decrease in total tumor volume for the experimental group as a result of these changes. Tables \ref{Tbl:ScnB_PC3_B} and \ref{Tbl:ScenB_DU_B} in Appendix \ref{AppendixA} list parental and resistant end-of-week tumor volume sizes for the PC3 and DU145 cell lines in Scenario A, respectively.}

    \item {Scenario A$^\prime$ involved a similar process, administering 2 Gy every day for five days as our control, and 1.75 Gy Monday-Thursday and 3 Gy on Friday as our experimental schedule. We remark that the control case is one of the standard clinical treatment schedules. This totals 10 Gy at the end of every week. Similar results to Scenario A were observed, with a steady decrease in end-of-week tumor volume values. This change resulted in a much larger difference between experimental and control values than Scenario A, however, and A$^\prime$ more clearly exhibits the benefits of a stronger Friday dosage. See tables \ref{TblB:PC3_A} and \ref{TblB:DU145_A} in Appendix \ref{AppendixA} for numerical results as well as Figure \ref{fig:table_5} for a graphical representation.}

%     \begin{figure}[htb!]
    
% \centerline{  
%         \includegraphics[width=4cm]{Scenerio A (Control).png}
%         \includegraphics[width=4cm]{DU145_ScenA_Control.png}
%         }
% \centerline{  
%         \includegraphics[width=4cm]{Scenerio A (Experimental).png}
%         \includegraphics[width=4cm]{DU145_ScenA_Exp.png}
%         }
%     \caption{Comparison of 2 Gy administered every weekday (top) vs. 1.75 Gy given from Monday-Thursday and 3 Gy every Friday (bottom). PC3 cell line left, DU145 cell line right. Only the first 2 weeks are depicted to show more detail. A decrease in values at the end of each week can be noted for both cell lines. A consistent dose will allow for more growth on the weekend, however a stronger Friday dose will create for a bigger dip in the graph, increasing overall effectiveness.}
%     \label{fig:scenA}
% \end{figure}
    
\end{itemize}

Scenarios {B} and {B$^\prime$} consist of a variation from scenarios A and {A$^\prime$}, in which a dose is administered every Monday, Wednesday and Friday  (skipping Tuesday and Thursday), still incorporating a stronger Friday dosage. {The motivations behind this were to incorporate some of the findings from earlier on in the paper to this dosage strategy as well, that being the effectiveness of increasing the time interval in between doses while maintaining a constant overall dosage.}

\begin{figure}[htb!]
\centerline{  
        \includegraphics[width=4.2cm]{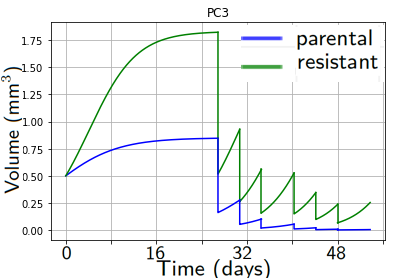}
        \includegraphics[width=4.2cm]{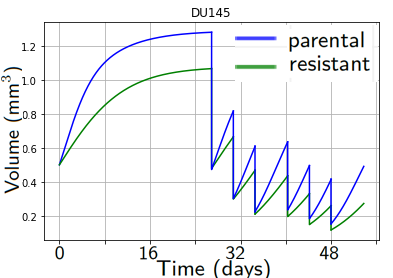}
        }
\centerline{  
        \includegraphics[width=4.2cm]{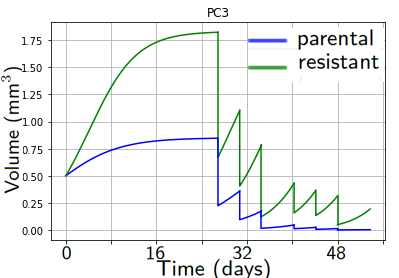}
        \includegraphics[width=4.2cm]{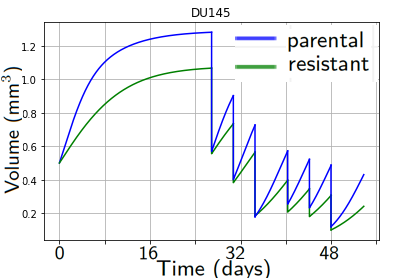}
        }
    \caption{{Scenario B} - A comparison of 3 Gy administered every Monday, Wednesday, and Friday {(Control, top)} vs. 2.5 Gy given on Monday and Wednesday, and 4 Gy every Friday {(Experimental, bottom)}, in a tumor {initiated} as a 1:1 mixture of {parental (blue)} and resistant {green} tumor cell populations. The results show the volumes for the individual {parental} and resistant populations from the PC3 cell line (left) and DU145 cell line (right). Only the first 2 weeks are depicted to show more detail. A modest decrease in values at the end of each week can be noted for both cell lines. A consistent dose will allow for more growth on the weekend, however a stronger Friday dose will yield a larger dip in the graph, increasing overall {dosing} effectiveness.}
    \label{fig:scenC}
\end{figure}

\begin{figure}[htb!]
\centerline{  
        \includegraphics[width=8cm]{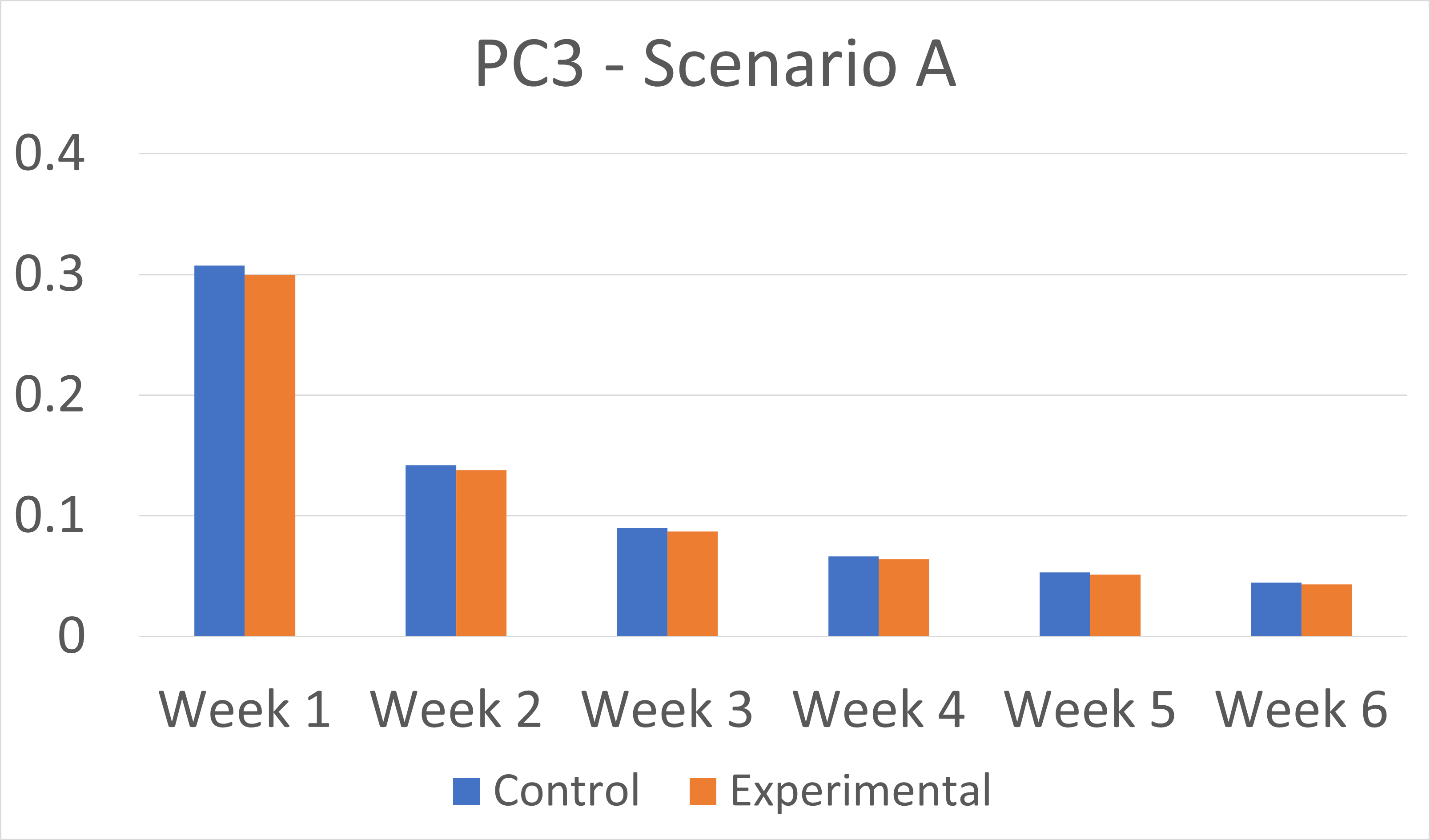}
        }
\centerline{  
        \includegraphics[width=8cm]{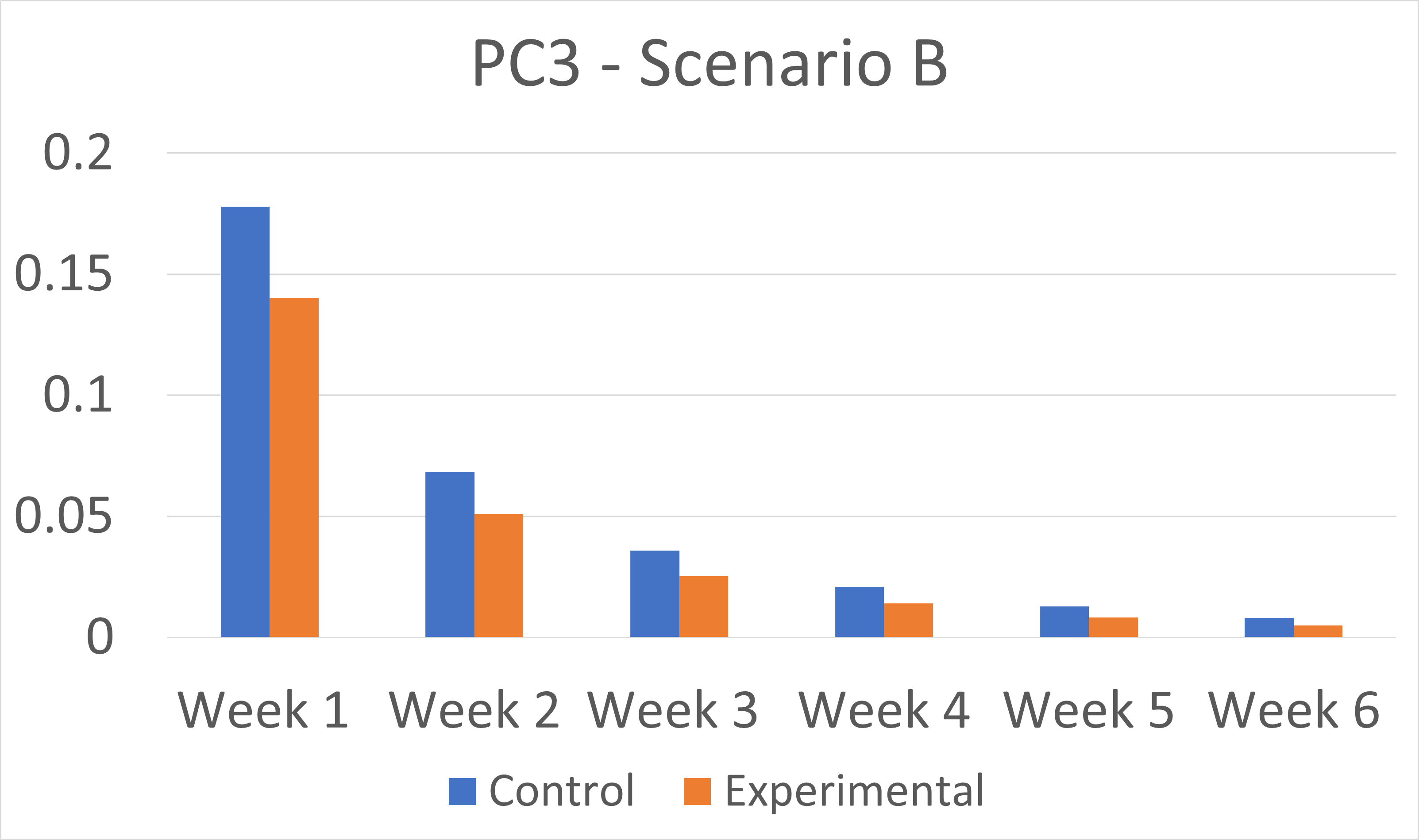}
        }
\caption{{PC3 cell line for Scenario A (top) and Scenario B (bottom). Each bar represents the total tumor volume at the end of each week for either the control population (blue) or the experimental population (orange). 9 Gy are administered every week for a period of 6 weeks for either scenario. Scenario A presents a slight decrease between experimental and control populations, but by combining the two strategies, Scenario B presents itself as a much more promising approach in decreasing overall tumor volume at the end of each week and across all 6 weeks of treatment.}}

    \label{fig:table5}
\end{figure}

\begin{figure}[htb!]
\centerline{ 
        \includegraphics[width=8cm]{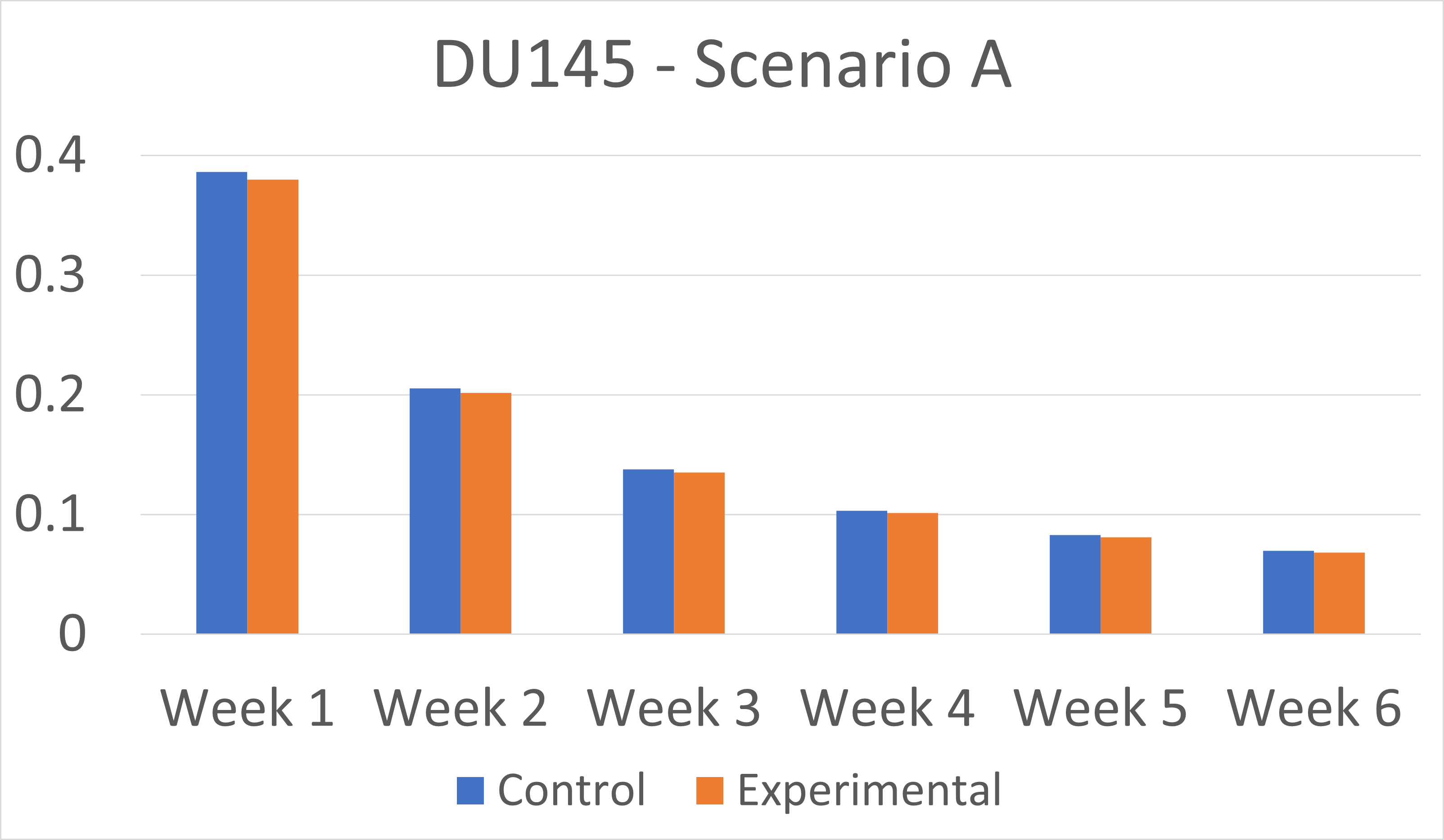}
        }
\centerline{  
        \includegraphics[width=8cm]{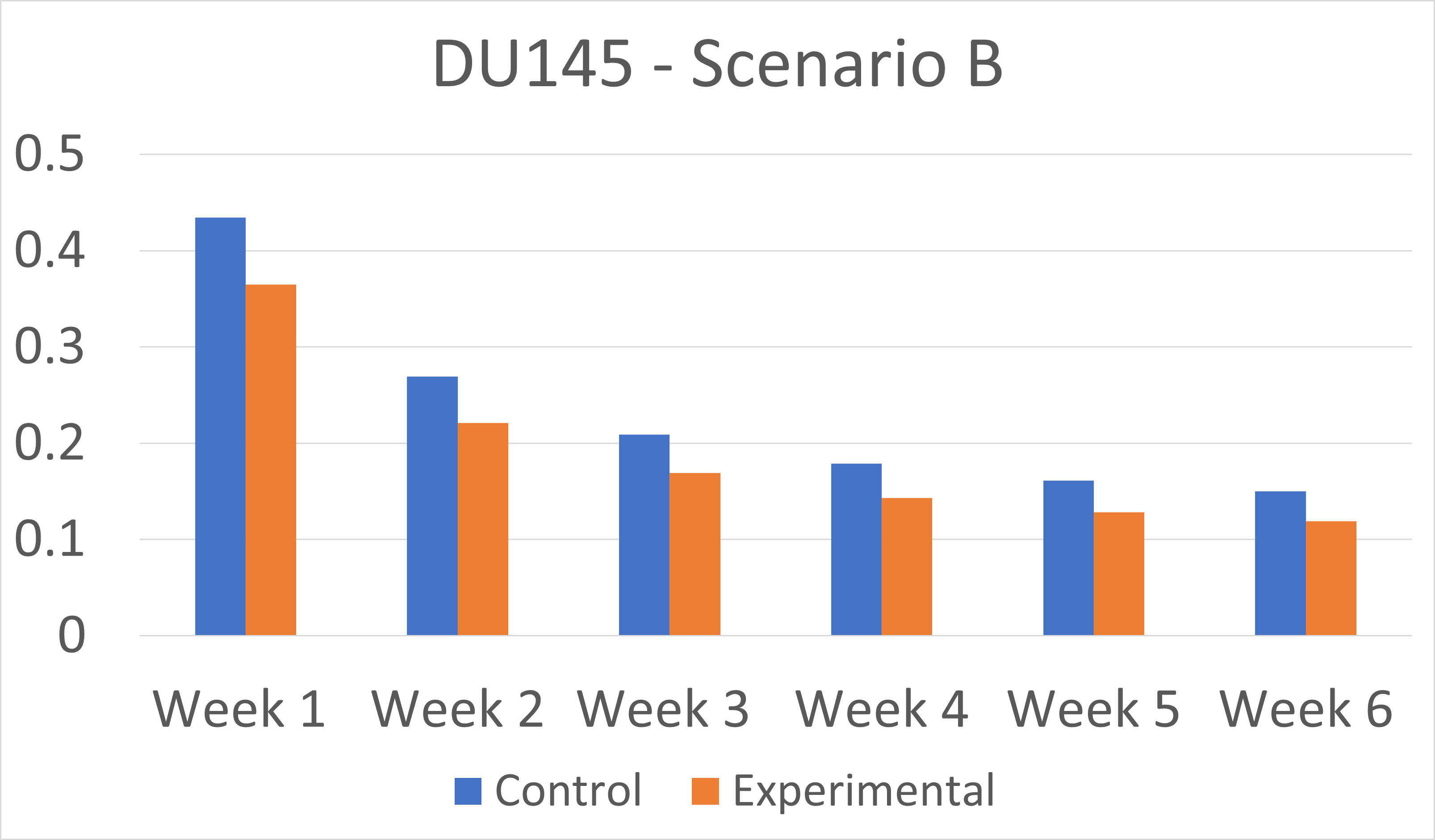}
        }
    \caption{{DU145 cell line for Scenario A (top) and Scenario B (bottom). Each bar represents total tumor volume at the end of each week. Control - blue, experimental - orange. 9 Gy are administered in a single week. DU145 presents with similar results as the PC3 cell line when following the same dosage protocol described in Scenario A, however as per the previous section, implementing increasing time intervals in between tumor fractionation is not as effective for the DU145 cell line as it is for the PC3 cell line, therefore it presents with worse overall results than Scenario A for DU145, confirming the ineffectiveness of hypofractionation on DU145.}}
    \label{fig:table6}
\end{figure}

\begin{itemize}
    \item Scenario {B} involved administering 3 Gy every Monday, Wednesday and Friday for our control. Our experimental regime included 2.5 Gy on Monday and Wednesday, and 4 Gy on Friday.  This totals 9 Gy at the end of each week. These values show a promising decrease in end-of-week tumor volume size when concentrating a greater dose on Fridays. Figure \ref{fig:scenC} depicts the first two weeks of this scenario in both the control and experimental groups for either cell line as a side-by-side comparison. {Figures \ref{fig:table5} and \ref{fig:table6} show the difference between control and experimental tumor volumes as a result of these changes in dosage for the PC3 and DU145 cell lines, respectively. See tables \ref{TblB:PC3_C} and \ref{TblB:DU145_C} in Appendix \ref{AppendixA} for end-of-week values.}

    \item Similarly to Scenario {B}, Scenario {B$^\prime$} involved administering {a certain dosage,} 2.5 Gy, every Monday, Wednesday and Friday for our control.  Our experimental schedule consists of 2 Gy on Monday and Wednesday, and 3.5 Gy on Friday.  This totals {7.5 Gy} at the end of the week each week. Similar results to Scenario {B} were observed, with a steady decrease in end-of-week tumor volume values. See tables \ref{Tb2:ScenB_PC3_D} and \ref{Tb2:ScenB_DU_D} in Appendix \ref{AppendixA} for numerical results as well as Figure \ref{fig:table_7} for a graphical representation. 
    
\end{itemize}

%{It is important to note that the difference between Scenarios A/{A$^\prime$} and Scenarios {B/B$^\prime$} the treatment schedule itself. Scenarios A and A$^\prime$ both offer treatment plans that would administer a consistent dosage everyday from Monday-Thursday, but a greater dosage on Friday. Scenarios B and B$^\prime$ both offer treatment plans that would administer the same dosage on Monday and Wednesday (skipping Tuesday and Thursday), and then a higher dosage on Friday.} 
Scenarios A and A$^\prime$ show us that by concentrating a higher dosage at the end of the week while maintaining a consistent weekly dosage, we end up with a smaller tumor volume at the end of each week for both cell lines. Scenarios B and B$^\prime$ combined this idea with the previously discussed idea of allowing for more time in between larger doses. {We reconfirm the effectiveness of concentrating a higher dosage on Friday compared to Monday and Wednesday in these scenarios. Smaller final tumor volumes are obtained in both cell lines. However, when we compare scenario A versus B, that is, smaller dosages every five days versus larger dosages on every other day, scenario B is more effective than scenario A only in the PC3 cell line, not in the DU145 cell line. This is consistent with the results obtained in section \ref{sec:simulation1}, hypofractionation is effective only in PC3 cell line, but not in DU145 cell line, even when combined with the stronger dosage on Friday strategy. We remark that increasing the total dosage using the same strategy always results in smaller tumor size. For instance, A$^\prime$ with total dosage 10 Gy is more effective than A with total dosage 9 Gy and B with total dosage 9 Gy is more effective than B$^\prime$ with total dosage 7.5 Gy. }

\section{Conclusion and Future Outlook}

In this study, we sought to study the effect of different radiation dosing strategies on the growth of prostate tumor spheroids consisting of a heterogeneous mixture of {parental} and radiation-resistant populations. Our results show that the administration of a higher dose with a longer period of time in between doses gives more promising results for the PC3 cell line, but not necessarily for the DU145 cell line. However, shifting dosage values so that more of it is administered on Friday and less during the week yields a moderate decrease in tumor volume at the end of each week for both cell lines. {Our results showed us that by simply shifting dosage administration values so that more of it is administered on Fridays, this will yield lower overall tumor volume outcomes the more you shift the dosage values, as seen in both cell lines in Scenarios A and B. Combining this shift in dosage values with hypofractionation strategies yields much more promising results in PC3 cell line, as noted in Scenarios B and B$^\prime$.} It is important to note that all tests were run in a simulated 1:1 ratio environment, and it would be insightful to look into other heterogeneous mixtures of {parental} and radiation resistant tumor populations. It is also important to note that the significant difference in results from the PC3 and DU145 cell lines shows us the impact of the diversity of the tumor population itself on radiation treatment. 

\begin{figure}[htb!]
\centerline{ 
        \includegraphics[width=8.5cm]{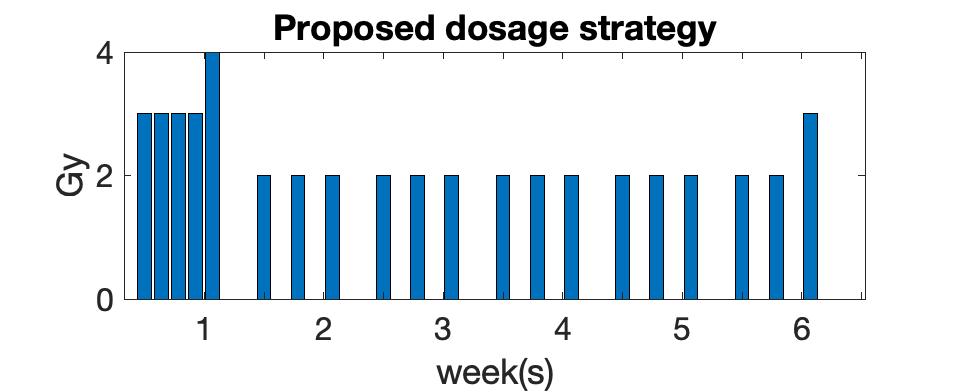}
        }
\centerline{ 
        \includegraphics[width=8cm]{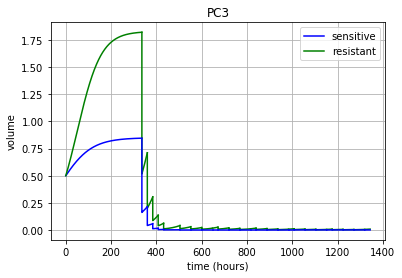}
        }
    \caption{Proposed dosage strategy (top) applied to a tumor {initiated} as a 1:1 mixture of parental and resistant PC3 cell line (bottom). In the first week, 3 Gy administered daily, with 4 Gy administered on the first Friday. Afterward a constant dose of 2 Gy is administered every Monday, Wednesday, and Friday for 5 weeks, with a final dose of 3 Gy administered at the end of the 6 week time period. The results show the volumes for the individual parental and resistant populations. {As detailed in Tables \ref{TblC1} and \ref{TblC2}, our proposed dosage strategy results in the smallest final tumor volume compared to all dosage plans tested in the previous sections and the standard clinical schedule.} }
    \label{fig:test1}
    
\end{figure}

We conclude from our results that we can improve the treatment for a 1:1 PC3 cell line mixture using a combination of both methods tested. Many different types of treatments are possible in this manner, but one such strategy is proposed in Figure \ref{fig:test1}. In this scheme, 3 Gy of radiation is administered daily for the first week, with a 4 Gy dose administered on the first Friday. Afterward a constant dose of 2 Gy is administered every Monday, Wednesday, and Friday for 5 weeks, with a final dose of 3 Gy administered at the end of the 6 week time period. No radiation is administered on any weekend. {This sample combination strategy assimilates both ideas described in this paper; such that when the tumor is at its largest we may give a stronger overall dose at the first week with a stronger dose on the first Friday, and then afterward apply hypofractionation strategies from weeks 2-6. Because a standard clinical radiotherapy procedure would involve applying 10 Gy per week, and in this scenario the tumor undergoes standard radiotherapy procedures for a 6-week time frame, then this proposal strategy follows that 60 Gy limit while providing a higher dose of radiation at the first week. While this greatly slows future growth from weeks 2-6 in the model (along with the effectively proven implemented hypofractionation strategies), the effects of increasing the radiation dosage in the first week by about 60\% at the outset may have adverse biological effects that would need to be observed in a laboratory setting.}

\begin{table*}[t] \centering
	\begin{tabular}{|c|c|c|c|} \hline 
	 & & {Standard Clinical Radiotherapy Schedule} & Proposed Dosage Strategy \\ \hline 
	 \multirow{3}{*}{Final Tumor Volume} & Parental & {1.0739 $\cdot 10^{-7}$ }& 1.0790 $\cdot 10^{-8}$\\ 
	 & Resistant & {0.030486} & 0.0074191 \\
	 & Total & {0.030486} & 0.0074191\\ \hline 
	 \multirow{3}{*}{Area Under Curve} & Parental & {19.910} & 6.1733 \\ & Resistant & {140.41} & 33.693\\ & Total & {160.32} & 39.866\\ \hline

\hline 
	\end{tabular}
	\caption{Comparison of values between {data collected from our simulation of the standard clinical radiotherapy schedule (a constant 2 Gy dose every weekday with a break on weekends)} and results from our proposed dosage strategy. {Both models used PC3 cell line parameters.} A clear decrease in final tumor volume and total area can be observed in both parental and resistant populations after implementation of the proposed dosage strategy.}
	\label{TblC1}
\end{table*}

\begin{table*}[t] \centering
	\begin{tabular}{|c|c|c|c|c|c|c|c|} \hline 
\multicolumn{2}{|c|}{Tumor Volume} & week 1 & week 2 & week 3 & week 4 & week 5 & week 6\\ \hline 
\multirow{3}{*}{ \begin{tabular}{c}
{Clinical}  \\ {Standard}
\end{tabular} } & {Parental} & {0.051106} & {0.0077640} & {0.0013167} & {0.00022713} & {3.9305 $\cdot 10^{-5}$} & {6.8229 $\cdot 10^{-6}$} \\ 
& {Resistant} & {0.15534} & {0.064823} & {0.034116} & {0.019652} & {0.011837} & {0.0073082}\\ 
 & {Total} & {0.20645} & {0.072587} & {0.035433} & {0.019879} & {0.011876} & {0.0073150}
\\ \hline
\multirow{3}{*}{Scenario A} & Parental & 0.044009 & 0.0063787 & 0.0010296 & 1.6888$\cdot 10^{-4}$ & 2.7777$\cdot 10^{-5}$  & 4.5829$\cdot 10^{-6}$  \\ & Resistant & 0.13129 & 0.052691 & 0.026704 & 0.014785 & 0.0085415 & 0.0050479 \\ 
 & Total & 0.17530 & 0.059070 & 0.027734 & 0.014954 & 0.0085693 & 0.0050525
 \\ \hline 
\multirow{3}{*}{ \begin{tabular}{c}
Proposed  \\ Strategy
\end{tabular} } & Parental & 0.00039753 & 3.4299$\cdot 10^{-5}$ & 2.9631$\cdot 10^{-6}$ & 2.5612$\cdot 10^{-7}$ & 2.2186$\cdot 10^{-8}$ & 3.6478$\cdot 10^{-9}$ \\ 
 & Resistant & 0.0097035 & 0.0061374 & 0.0039301 & 0.0025354 & 0.0016432 & 0.0017628 \\ 
 & Total & 0.010101 & 0.0061717 & 0.0039331 & 0.0025357 & 0.0016432 & 0.0017628
\\ \hline 
	\end{tabular}
	\caption{Comparison of values between {the standard clinical radiotherapy model,)} results referenced in Table \ref{TblB:PC3_A}, and our proposed dosage strategy. Scenario A involved administering 1.75 Gy every day from Monday through Thursday and then 3 Gy every Friday, as well as withholding any radiation on weekends. We mirrored this scheme more effectively in our proposed strategy, as exhibited by the decrease in end-of-week values for all six weeks. {Our proposed strategy proved to be most effective.}}
	\label{TblC2}
\end{table*}

{In essence, this proposal offers a starting point to what may be a simulation-based approach to radiotherapy that would need to be balanced with the biological limitations of living cells. A more balanced approach would only be possible after testing various simulation-based radiotherapy dosage strategies in a laboratory setting, and then possibly using a feedback approach to integrate the results of each tested strategy into a simulation that can better optimize each strategy with the use of more data.} 

When it comes to a 1:1 PC3 cell line mixture, comparison of this dosage strategy to our previous findings shows us that implementation of both the usage of constant dosage strategies with adaptive dose levels, as well as a stronger dose on Fridays, proves to be more effective than using just one of the two strategies by itself. Table \ref{TblC1} compares values between our proposed dosage strategy and results from Table \ref{TblA:PC3}, and Table \ref{TblC2} compares values between Scenario A, defined in Table \ref{TblB:PC3_A}, and our proposed dosage plan. Both data sets show a clear drop in tumor volumes, proving the effectiveness of combining both strategies together. 

{Clinicians may be able to propose varying dosage strategies prior to the treatment cycle depending on the cell mixture, cell line, size, and growth rate of the tumor in order to fit the needs of each individual patient. This proposal would be most effective in theory on this particular cell line (PC3) within a 1:1 mixture of parental to radio-resistant cells. Due to the particularity of each tumor in any individual patient, the importance of math modeling in clinical radiotherapy remains well-grounded, as ODE models can be used to predict the most effective course of treatment in any particular case, as opposed to following standard procedure that does not take into account all of the variables involved in each individual tumor population, such as tumor heterogeneity. Our hope is that in the future, with the help of ever-advancing medical technology, patients' tumors may be able to be paramaterized in a similar manner to this study, and then those parameters can be modeled using ODE's that can extrapolate the approximate growth behavior of the tumor. Using this extrapolation, the best possible dosage strategy can be approximated using much more advanced optimization strategies than those implemented here, but also one that builds upon the framework proposed earlier in this paper.} 

{Of note to point out is that all experiments in this simulation study were done \textit{in silico}, and although parameters were derived from experiments, the natural world contains many other complex parameters that cannot yet be replicated within a computer simulation. Therefore, further means of study and real-life investigations are necessary before we can apply the findings of this study to a clinical application. That being said, o}ur work underscores the potential of adaptive cancer treatments, which is of recent interest in the field. In particular, this concept can be applied to those patients that have heterogeneous mixtures of cancer, with  both temporal and spatial variability in their cancer microenvironment and even therapy-induced perturbations.

In our future work, we propose to do further analysis into different prostate {cancer} cell lines, such as the DU145 cell line to examine the efficacy of adaptive dosages. {While the PC3 cell line showed promising results using the hypofractionation strategy, the DU145 cell line did not. This could be due to various factors, including the differences in the proliferation rate and interaction types of the populations. We aim to further identify the key parameters and underlying mechanisms that yield such differences to identify the cancer types and patients that hypofractionation strategy should be applied to. Our final goal is} to develop strategies for cancer cell lines with different interaction properties to improve treatment outcome. {Comparing other radiotherapy models with our current choice, the linear-quadratic model with instant response, will be our future work as well. It will be interesting to validate our results across different radiotherapy models including a logistic type of radiotherapy model \cite{Prokopiou2015,Poleszczuk2018}, dynamic carrying capacity \cite{Mohammad2021}, and delayed response from necrotic population \cite{Cho2020}.} 
In the future, we also hope to do prospective experiments to examine the efficacy of proposed dosage strategies, {in addition to studying other initial mixture ratios}.

%It is imperative to clarify that this was a simulation-based study that has yet to be verified experimentally and clinically. Our study using ordinary differential equations  simplifies the complex reality of the tumor, and this may lead to disjunction between our results and reality. We hope to study the factors that close to gap between the simulation and reality in the future.

%However, our data show us that there can be many different approaches to implementing dosage strategies for radiation treatment of cancer cell lines, and our proposed dosage scheme is just one approach out of many different combinations of strategies that can be taken in the radiation treatment of heterogeneous tumors.

%\parbox{1\linewidth}{\vspace{1cm} something something}

%%%%%%%%%%%%%%%%%%%%%%%%%%%%%%%%%%%%%%%%
%%%%%%%%%%%%%%%%%%%%%%%%%%%%%%%%%%%%%%%%

\section*{Code Availability}

Code is available at https://github.com/josedualv/Effective-dose-fractionation-schemes-of-radiotherapy-on-prostate-cancer. 

\section*{Author Contributions}

% This is a required section for the journal.
% Use this section to indicate each author’s individual contribution to the finished manuscript.
% You might consider the \href{https://en.wikipedia.org/wiki/Taxonomy_(general)#Taxonomies_in_research_publishing}{Contributor Roles Taxonomy} for a list of possible contributor roles.

Conceptualization, H.C.; 
Methodology, J.A., H.C.; 
Software, J.A., H.C., K.S.; 
Data curation, P.K.;
Formal Analysis, J.A.; 
Investigation, J.A.; 
Writing – Original Draft, J.A.; 
Writing – Review \& Editing, J.A., H.C., P.K., K.S.; 
Visualization, J.A., H.C.;

%%%%%%%%%%%%%%%%%%%%%%%%%%%%%%%%%%%%%%%%
%%%%%%%%%%%%%%%%%%%%%%%%%%%%%%%%%%%%%%%%
% \printbibliography

\bibliographystyle{spmpsci.bst}
\bibliography{reference_Radiotherapy.bib}

% \begin{thebibliography}{9}

% \bibitem{Au1} Author, A. (1993). \emph{Authoritative book on the subject.} City: Publisher.

% \bibitem{OWAu} O'Wittahl, K.\,N., \& Author, A. (1999). A study of something interesting. \emph{Prestigious Journal, 180}(2), 293--316.

% \bibitem{Au2} Author, A. (2002). Cute paper that expands on previous results. \emph{Small Journal, 18}, 108--116.

% \end{thebibliography}

\appendix 
\section{{Tables and Bar Graphs for Scenarios A, B, A$^\prime$, and B$^\prime$}}\label{AppendixA}

\setcounter{table}{0}
\setcounter{figure}{0}
\renewcommand{\thefigure}{A\arabic{table}}
\renewcommand{\thefigure}{A\arabic{figure}}

    \begin{figure}[htb!]
\centerline{ \ 
        \includegraphics[width=8cm]{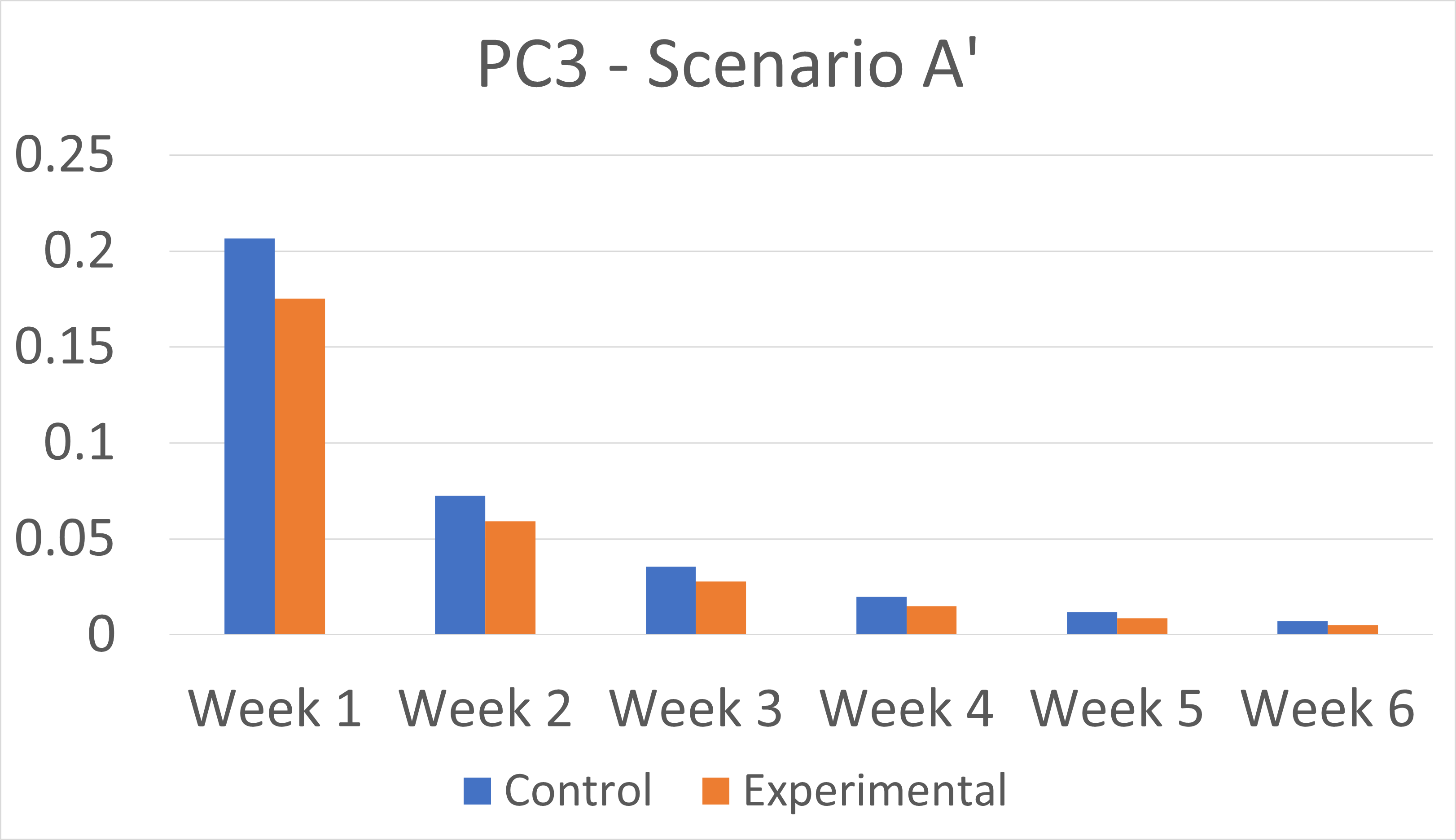}
        }
\centerline{ \ 
        \includegraphics[width=8cm]{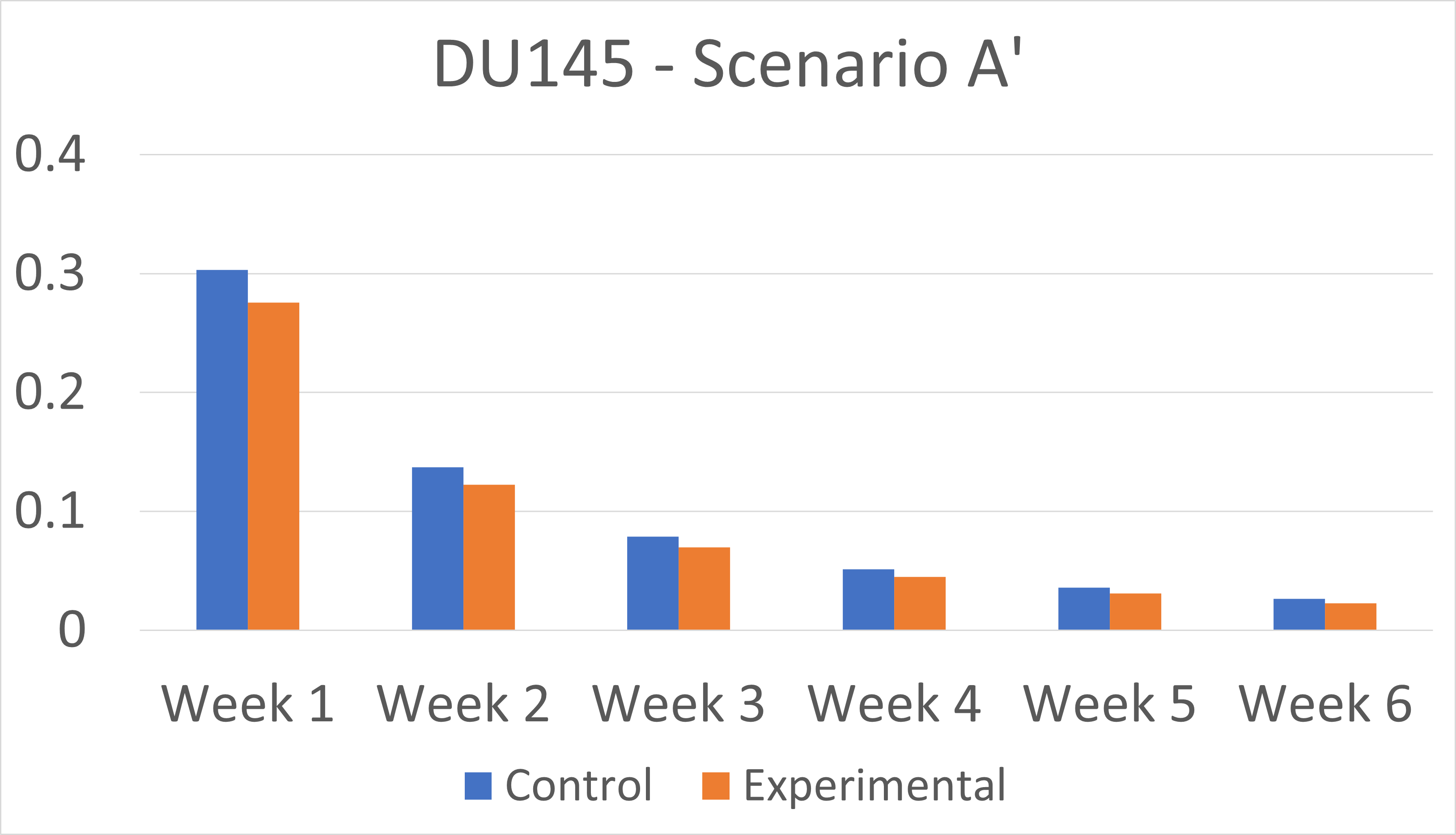}
        }
\caption{{End-of-week tumor volumes in graphical format for Scenario A$^\prime$ on the PC3 cell line (top) and on the DU145 cell line (bottom). Scenario A$^\prime$ shows the effectiveness of a higher Friday dosage without the implementation of hypofractionation strategies on either cell line (as in Scenarios B and B$^\prime$). While weekly dosage is higher than that of Scenarios A and B (10 Gy vs 9 Gy), this value is consistent between control and experimental simulations within this scenario, meaning the difference in end-of-week tumor volumes seen between the control and experimental simulations is entirely due to a stronger Friday dosage.}}
    \label{fig:table_5}
\end{figure}

    \begin{figure}[htb!]
\centerline{ \ 
        \includegraphics[width=8cm]{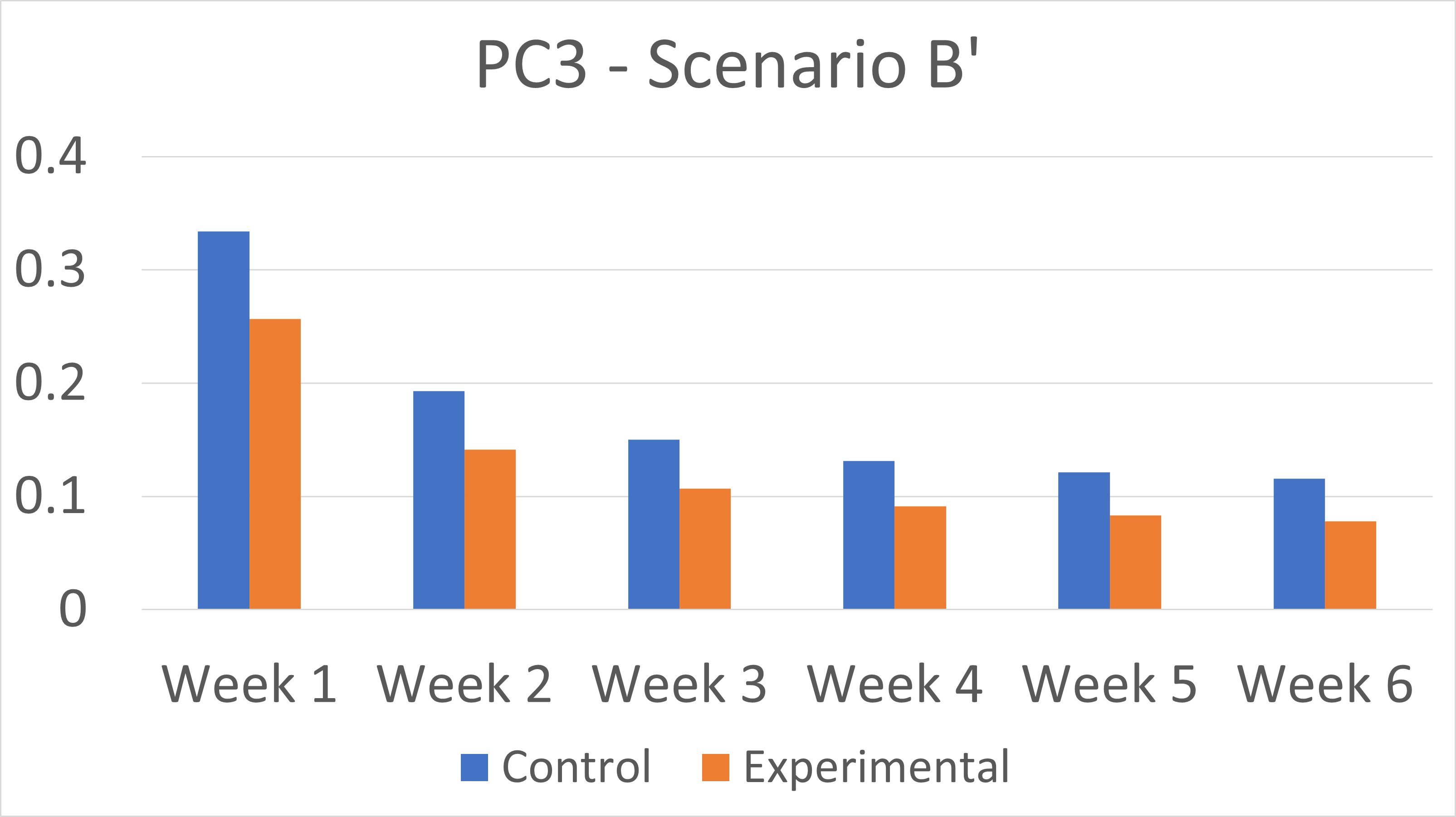}
        }
\centerline{ \ 
        \includegraphics[width=8cm]{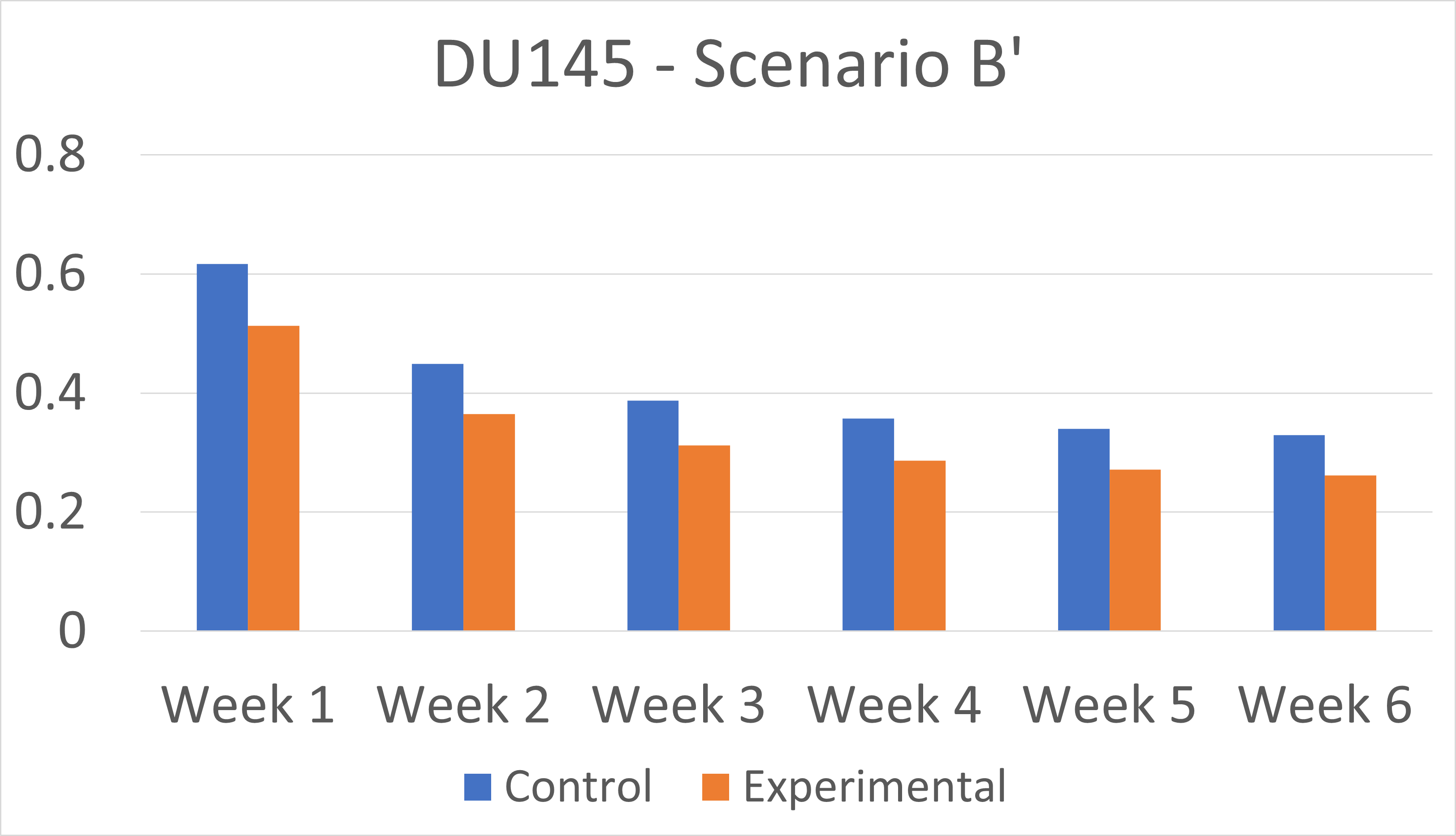}
        }
    \caption{{End-of-week tumor volumes in graphical format for Scenario B$^\prime$ on the PC3 cell line (top) and on the DU145 cell line (bottom). This scenario involves administration of only 7.5 Gy every week, and implements a stronger Friday dosage as well as hypofractionation strategies discussed earlier. While a stronger Friday dosage is still effective for either cell line, hypofractionation strategies remain minimally effective to the overall tumor volume of DU145 cell line.}}
    \label{fig:table_7}
\end{figure}

\begin{table*}[!htb] \centering
{
	\begin{tabular}{|c|c|c|c|c|c|c|c|} \hline 
\multicolumn{2}{|c|}{Tumor Volume} & week 1 & week 2 & week 3 & week 4 & week 5 & week 6 \\ \hline 
\multirow{3}{*}{Control} & Parental & 0.084901 & 0.021027 & 0.0061529 & 0.0018777 & 0.00058006 & 0.00017988
\\ & Resistant & 0.22261 & 0.12080 & 0.083701 & 0.064383 & 0.052479 & 0.044384 \\ 
 & Total & 0.30751 & 0.14183 & 0.089854 & 0.066261 & 0.053059 & 0.044564\\ \hline 
\multirow{3}{*}{Experimental} & Parental & 0.082921 & 0.020487 & 0.0059828 & 0.0018221 & 0.00056178 & 0.00017386\\ & Resistant & 0.21667 & 0.11725 & 0.081114 & 0.062311 & 0.050729 & 0.042854 \\ 
 & Total & 0.29959 & 0.13774 & 0.087097 & 0.064133 & 0.051291 & 0.043028\\ \hline 
	\end{tabular}
	\caption{PC3 cell line for Scenario A. Values represent tumor volume at the end of each week. Control: 1.8 Gy administered Monday through Friday; Experimental: 1.75 Gy administered Monday through Thursday, 2 Gy administered on Friday. 9 Gy administered in a single week.}
	\label{Tbl:ScnB_PC3_B}}
\end{table*}

\begin{table*}[!htb] \centering
{
	\begin{tabular}{|c|c|c|c|c|c|c|c|} \hline 
\multicolumn{2}{|c|}{Tumor Volume} & week 1 & week 2 & week 3 & week 4 & week 5 & week 6 \\ \hline 
\multirow{3}{*}{Control} & Parental & 0.15188 & 0.066600 & 0.036114 & 0.021379 & 0.013240 & 0.0084134
\\ & Resistant & 0.23423 & 0.13887 & 0.10169 & 0.081892 & 0.069616 & 0.061283 \\ 
 & Total & 0.38611 & 0.20547 & 0.13780 & 0.10327 & 0.082856 & 0.069696\\ \hline 
\multirow{3}{*}{Experimental} & Parental & 0.14927 & 0.065328 & 0.035387 & 0.020933 & 0.012954 & 0.0082259
\\ & Resistant & 0.23038 & 0.13631 & 0.099727 & 0.080260 & 0.068197 & 0.060011 \\ 
 & Total & 0.37965 & 0.201638 & 0.13511 & 0.10119 & 0.081151 & 0.068237 \\ \hline 
	\end{tabular}
	\caption{DU145 cell line for Scenario A. Values represent tumor volume at the end of each week. Control: 1.8 Gy administered Monday through Friday; Experimental: 1.75 Gy administered Monday through Thursday, 2 Gy administered on Friday. 9 Gy administered in a single week.}
	\label{Tbl:ScenB_DU_B}}
\end{table*}

\begin{table*}[t] \centering
{
	\begin{tabular}{|c|c|c|c|c|c|c|c|} \hline 
\multicolumn{2}{|c|}{Tumor Volume} & week 1 & week 2 & week 3 & week 4 & week 5 & week 6 \\ \hline 
\multirow{3}{*}{Control} & Parental & 0.019411 & 0.0015771 & 0.00013555 & 1.1705$\cdot 10^{-5}$ & 1.0112$\cdot 10^{-6}$ & 8.7391$\cdot 10^{-8}$
\\ & Resistant & 0.15839 & 0.066878 & 0.035654 & 0.020827 & 0.012734 & 0.0079866 \\ 
 & Total & 0.17780 & 0.068455 & 0.035790 & 0.020839 & 0.012735 & 0.0079867
\\ \hline 
\multirow{3}{*}{Experimental} & Parental & 0.016345 & 0.0012511 & 0.00010121 & 8.2218$\cdot 10^{-6}$ & 6.6823$\cdot 10^{-7}$ & 5.4331$\cdot 10^{-8}$ \\ & Resistant & 0.12392 & 0.049846 & 0.025389 & 0.014144 & 0.0082270 & 0.0048978 \\ 
 & Total & 0.14027 & 0.051097 & 0.025490 & 0.014152 & 0.0082277 & 0.0048979
 \\ \hline 
	\end{tabular}
	\caption{PC3 cell line for Scenario B. Values represent tumor volume at the end of each week. Control: 3 Gy administered Monday, Wednesday, and Friday; Experimental: 2.5 Gy administered Monday and Wednesday, 4 Gy on Friday. 9 Gy administered in a single week.}
	\label{TblB:PC3_C}}
\end{table*}

\begin{table*}[t] \centering
{
	\begin{tabular}{|c|c|c|c|c|c|c|c|} \hline 
\multicolumn{2}{|c|}{Tumor Volume} & week 1 & week 2 & week 3 & week 4 & week 5 & week 6 \\ \hline 
\multirow{3}{*}{Control} & Parental & 0.22442 & 0.15275 & 0.12874 & 0.11798 & 0.11256 & 0.10967
\\ & Resistant & 0.20989 & 0.11644 & 0.080110 & 0.060718 & 0.048637 & 0.040383 \\ 
 & Total & 0.43431 & 0.26919 & 0.20881 & 0.17870 & 0.16120 & 0.15005
 \\ \hline 
\multirow{3}{*}{Experimental} & Parental & 0.18009 & 0.12056 & 0.10093 & 0.092219 & 0.087874 & 0.085586
\\ & Resistant & 0.18451 & 0.10023 & 0.068033 & 0.050978 & 0.040404 & 0.033206 \\ 
 & Total & 0.36460 & 0.22079 & 0.16896 & 0.14320 & 0.12828 & 0.11879
 \\ \hline 
	\end{tabular}
	\caption{{DU145 cell line for Scenario B. Values represent tumor volume at the end of each week. Control: 3 Gy administered Monday, Wednesday, and Friday; Experimental: 2.5 Gy administered Monday and Wednesday, 4 Gy on Friday. 9 Gy administered in a single week.}}
	\label{TblB:DU145_C}}
\end{table*}

\begin{table*}[t] \centering
{
	\begin{tabular}{|c|c|c|c|c|c|c|c|} \hline 
\multicolumn{2}{|c|}{Tumor Volume} & week 1 & week 2 & week 3 & week 4 & week 5 & week 6 \\ \hline 
\multirow{2}{*}{Control} & Parental & 0.051106 & 0.0077640 & 0.0013167 & 0.00022712 & 0.000039305 & 0.0000068229 \\ & Resistant & 0.15534 & 0.064823 & 0.034116 & 0.019652 & 0.011837 & 0.0073082\\ & Total & 0.20645 & 0.072587 & 0.035433 & 0.019879 & 0.011876 & 0.0073150
\\ \hline 
\multirow{2}{*}{Experimental} & Parental & 0.044009 & 0.0063787 & 0.0010296 & 0.00016888 & 0.000027777 & 0.0000045829 \\ & Resistant & 0.13129 & 0.052691 & 0.026704 & 0.014785 & 0.0085415 & 0.0050479\\ 
 & Total & 0.17530 & 0.059070 & 0.027734 & 0.014954 & 0.0085693 & 0.0050525
 \\ \hline 
	\end{tabular}
	\caption{PC3 cell line for Scenario A$^\prime$. Control: 2 Gy administered Monday through Friday; Experimental: 1.75 Gy administered Monday through Thursday, 3 Gy administered on Friday. 10 Gy administered in a single week.}
	\label{TblB:PC3_A}}
\end{table*}

\begin{table*}[t] \centering
{
	\begin{tabular}{|c|c|c|c|c|c|c|c|} \hline 
\multicolumn{2}{|c|}{Tumor Volume} & week 1 & week 2 & week 3 & week 4 & week 5 & week 6 \\ \hline 
\multirow{2}{*}{Control} & Parental & 0.11442 & 0.040030 & 0.017021 & 0.0077624 & 0.0036484 & 0.0017389 \\ & Resistant & 0.18871 & 0.096959 & 0.061894 & 0.043493 & 0.032255 & 0.024753\\ & Total & 0.30313 & 0.13699 & 0.078915 & 0.051255 & 0.035903 & 0.026492
\\ \hline 
\multirow{2}{*}{Experimental} & Parental & 0.10364 & 0.035558 & 0.014866 & 0.0066652 & 0.0030785 & 0.0014413\\ & Resistant & 0.17204 & 0.086988 & 0.054900 & 0.038185 & 0.028037 & 0.021301\\ & Total & 0.27568 & 0.12255 & 0.069766 & 0.044850 & 0.031116 & 0.022742
\\ \hline 
	\end{tabular}
	\caption{DU145 cell line for Scenario A$^\prime$. Control: 2 Gy administered Monday through Friday; Experimental: 1.75 Gy administered Monday through Thursday, 3 Gy administered on Friday. 10 Gy administered in a single week.}
	\label{TblB:DU145_A}}
\end{table*}

\begin{table*}[!htb] \centering
	\begin{tabular}{|c|c|c|c|c|c|c|c|} \hline 
\multicolumn{2}{|c|}{Tumor Volume} & week 1 & week 2 & week 3 & week 4 & week 5 & week 6 \\ \hline 
\multirow{2}{*}{Control} & Parental & 0.046328 & 0.0090815 & 0.0020226 & 0.00046238 & 0.00010633 & 0.000024483
\\ & Resistant & 0.28760 & 0.18373 & 0.14793 & 0.13083 & 0.12139 & 0.11577
\\ & Total & 0.33393 & 0.19281 & 0.14995 & 0.13129 & 0.12150 & 0.11579
\\ \hline 
\multirow{2}{*}{Experimental} & Parental & 0.037598 & 0.0069530 & 0.0014604 & 0.00031438 & 0.000068035 & 0.000014740 \\ & Resistant & 0.21896 & 0.13435 & 0.10526 & 0.091107 & 0.083055 & 0.078059\\ & Total & 0.25656 & 0.14130 & 0.10672 & 0.091421 & 0.083123 & 0.078074
\\ \hline 
	\end{tabular}
	\caption{{PC3 cell line for Scenario B$^\prime$. Values represent tumor volume at the end of each week.} Control: 2.5 Gy administered Monday, Wednesday, and Friday; Experimental: 2 Gy administered Monday and Wednesday, 3.5 Gy on Friday. {7.5} Gy administered in a single week.}
	\label{Tb2:ScenB_PC3_D}
\end{table*}

\begin{table*}[!htb] \centering
	\begin{tabular}{|c|c|c|c|c|c|c|c|} \hline 
\multicolumn{2}{|c|}{Tumor Volume} & week 1 & week 2 & week 3 & week 4 & week 5 & week 6 \\ \hline 
\multirow{2}{*}{Control} & Parental & 0.32275 & 0.24864 & 0.22279 & 0.21006 & 0.20257 & 0.19766 \\ & Resistant & 0.29400 & 0.20008 & 0.16489 & 0.14745 & 0.13762 & 0.13169\\ & Total & 0.61675 & 0.44872 & 0.38768 & 0.35751 & 0.34019 & 0.32935
\\ \hline 
\multirow{2}{*}{Experimental} & Parental & 0.25674 & 0.19434 & 0.17308 & 0.16272 & 0.15668 & 0.15274 \\ & Resistant & 0.25608 & 0.17060 & 0.13906 & 0.12344 & 0.11460 & 0.10922\\ & Total & 0.51282 & 0.36494 & 0.31214 & 0.28616 & 0.27128 & 0.26196
\\ \hline 
	\end{tabular}
	\caption{{DU145 cell line for Scenario B$^\prime$. Values represent tumor volume at the end of each week.} Control: 2.5 Gy administered Monday, Wednesday, and Friday; Experimental: 2 Gy administered Monday and Wednesday, 3.5 Gy on Friday. {7.5} Gy administered in a single week.}
	\label{Tb2:ScenB_DU_D}
\end{table*}

\end{document}